\documentclass[%
 reprint,
nofootinbib,
 amsmath,amssymb,
 aps,
]{revtex4-2}

\usepackage{graphicx}
\usepackage{dcolumn}
\usepackage{bm}
\usepackage{hyperref}

\usepackage{lipsum}

\def\al{\alpha}
\def\be{\beta}
\def\ga{\gamma}
\def\de{\delta}
\def\ep{\epsilon}
\def\ze{\zeta}
\def\et{\eta}
\def\th{\theta}
\def\ka{\kappa}
\def\la{\lambda}
\def\rh{\rho}

\def\si{\sigma}

\def\ph{\phi}

\def\Ga{\Gamma}

\def\La{\Lambda}
\def\Si{\Sigma}

\def\Ph{\Phi}
\def\Ps{\Psi}
\def\Om{\Omega}

\def\cL{{\mathcal L}}
\def\cR{{\mathcal R}}
\def\cD{{\mathcal D}}
\def\cH{{\mathcal H}}
\def\cG{{\mathcal G}}

\def\lie{{\cL_{\bf n}}}

\def\sl{{  s_{\bar{0}\bar{0}}  }} 
\def\dsl{{ \dot{s}_{\bar{0}\bar{0}} }}

\def\mn{{\mu\nu}}
\def\ab{{\al\be}}
\def\gd{{\ga\de}}

\def\abgd{{\al\be\ga\de}}

\def\sb{\overline{s}}
\def\st{\tilde{s}}

\def\pt#1{\phantom{#1}}
\def\prt{\partial}
\def\3g#1#2#3{^{(3)}\Ga^{#1}_{\pt{#1}#2#3}}

\def\fr#1#2{{{#1} \over {#2}}}

\def\etal {{\it et al.}}

\newcommand{\beq}{\begin{equation}}
\newcommand{\eeq}{\end{equation}}
\newcommand{\bea}{\begin{eqnarray}}
\newcommand{\eea}{\end{eqnarray}}
\newcommand{\rf}[1]{(\ref{#1})}

\begin{document}


\title{A 3+1 Formulation of the Standard-Model Extension Gravity Sector}

\author{Kellie O'Neal-Ault}
\email{aultk@my.erau.edu}
\author{Quentin G. Bailey}
\email{baileyq@erau.edu}
\affiliation{Embry-Riddle Aeronautical University, 3700 Willow Creek Road, Prescott, AZ, 86301, USA}
\author{Nils A. Nilsson}
\affiliation{National Centre for Nuclear Research, ul. Pasteura 7, 00-293, Warsaw, Poland}
\email{albin.nilsson@ncbj.gov.pl}





\date{\today}

\begin{abstract}
We present a 3+1 formulation of the effective field theory framework called the
Standard-Model Extension in the gravitational sector. 
The explicit local Lorentz and diffeomorphism symmetry breaking 
assumption is adopted and we perform a Dirac-Hamiltonian analysis. 
We show that the structure of the dynamics presents significant differences 
from General Relativity and other modified gravity models.  
We explore Hamilton's equations for some special choices of the coefficients.
Our main application is cosmology and we present the modified Friedmann equations 
for this case.
The results show some intriguing modifications to standard cosmology.
In addition, 
we compare our results to existing frameworks and models 
and we comment on the potential impact to other areas 
of gravitational theory and phenomenology.

\end{abstract}

\maketitle

\section{Introduction}

It is generally expected that General Relativity (GR) 
and the Standard Model (SM) of particle physics are not the ultimate descriptions of Nature, 
but rather low-energy effective field theories 
which accurately describe physics at energy scales available to us. 
This point of view is motivated by the expectation that there exists 
a single unified theory encompassing 
all the known fundamental interactions. 
This implies the existence of a renormalizable quantum theory of gravity 
which has GR as its low-energy limit. 
GR, 
being an effective field theory, 
is then expected to hold up to some ultraviolet (UV) cutoff scale, 
normally taken to be the Planck energy, 
$E_{\rm Pl} \approx 10^{19} GeV$. 
Any theory attempting to bridge GR and SM should, 
on dimensional grounds, 
contain all the characteristic constants of the constituent theories. 
As $E_{\rm Pl}$ represents the UV cutoff scale of GR, 
new physics should appear close to this energy, 
and a promising avenue to find new physics is to search for deviations 
from fundamental principles of GR.

Local Lorentz invariance is one of the fundamental symmetries of relativity 
as well as particle physics; 
stating that any local experiment 
is independent of both orientation 
and velocity of both the experiment and observer, 
and it is a key ingredient of GR. 
As such, precision tests of local Lorentz symmetry are an excellent way to test for new physics \cite{ks89kp95,review}. 

The Standard-Model Extension (SME) is a general effective field theory framework 
for testing Lorentz and CPT symmetries \cite{sme1,k04}. 
It has become a standard framework for constraining Lorentz violation 
in a systematic way (for a list of all current measurements, see Ref.\ \cite{datatables}). 
The SME contains GR and the Standard Model of particle physics, 
as well as generic Lorentz-violating terms up to arbitrary order. 
The terms are constructed by contracting operators built from known fields with 
coefficients for Lorentz violation, 
the latter of which control the degree of symmetry breaking
and can be constrained by experiments. 

In principle the SME contains an arbitrary number of terms, 
but is frequently truncated at low order 
in mass dimension of the field operators used. 
A much-studied truncation is called the \emph{minimal} SME 
and contains operators of mass dimension $3$ or $4$.

Whereas many limits have to-date been set in the matter sector of the SME, 
gravitational-sector coefficients have also been constrained.
These include test with short-range gravity tests \cite{short},
gravimeters \cite{gravi},
solar-system tests \cite{ss,llr}, 
pulsars \cite{pulsars}, 
gravitational waves \cite{gwtests},
and others \cite{othergrav}. 

Much of the theoretical phenomenology
that experiments and observations have used 
is based on weak-field gravity analysis \cite{bk06,wfanalysis,bkx15,km16,km18}.
So-called ``exact" results beyond this regime in the SME gravity sector 
have just begun to be explored \cite{bonder15,nl,b16,nab19,b19,kl20}. 
The aim of this work is in part to extend results to situations 
where weak gravitational fields cannot be assumed, 
for example in cosmology.
Furthermore, 
we begin a study of the 3+1 formulation of this framework, 
which allows for a Dirac-Hamiltonian analysis \cite{dirac}.
Note that this type of analysis has been performed for vector and 
tensor models of spontaneous Lorentz violation \cite{bgpv08,h14,ms1,ms2}
and other related models \cite{dj11}, 
but as of yet, 
has not been attempted for the SME and we seek to fill this gap in this work.
Primarily we shall adopt the explicit symmetry breaking scenario, 
though some of our results can be extended to spontaneous symmetry breaking.
Ultimately, 
we aim to push the application of the SME framework in a new direction 
in order to explore more broadly the consequences of Lorentz violation in gravity.

The SME as a framework for testing Lorentz symmetry 
naturally contains specific models of Lorentz violation as subsets.
Much work in the literature has involved the study of such models,
particularly in the gravity context 
\cite{ks89bb, ms09, models, bk05, bkx08, kp09, ms19-1}.
The connection between the coefficients for Lorentz violation in the SME
and proposed models in the literature has been established for 
some quantum gravity approaches \cite{gp99,km09}, 
massive gravity models \cite{bbw19}, 
noncommutative geometry \cite{nc} as well as vector and tensor models
of spontaneous Lorentz symmetry breaking.
In this paper, 
we use our results to match to yet another model which
involved explicit Lorentz breaking.

The paper is organized as follows: 
in Section~\ref{sec:SME} we give an overview of the key features of the SME. 
The details of the 3+1 decomposition are presented in Section~\ref{sec:ADM}, 
starting with a geometric overview followed by the discussion of 
the SME action terms. 
In Section~\ref{sec:ham} we perform a Hamiltonian analysis, 
starting with general features and we then focus on
two special cases.
As an application of the results, 
we study cosmological solutions in Section \ref{sec:cos}.
We connect our results to existing frameworks and models
in Section \ref{sec:match}.
Finally we discuss our results and conclusions in Section~\ref{sec:CON}, 
along with remarks on future work.

Notational conventions in this paper match prior work as much as possible \cite{k04,bk06}.  
Greek letters are used for spacetime indices and latin letters $i,j,k,...$ for spatial indices.
For local Lorentz frame (vierbein) indices we use the latin letters $a,b,c...$ when needed.
The metric $g_\mn$ signature matches the standard GR choice $(-+++)$ 
and we use units where $\hbar = c = 1$.
One important notational difference in this work is that we use here $\nabla_\mu$
for the spacetime covariant derivative, 
reserving $\cD_\mu$ for covariant derivatives defined 
on constant-time spatial hypersurfaces. 

\section{Basic Framework}\label{sec:SME}

In Riemann spacetimes, 
the lowest order terms in the GR + SME gravitational Lagrange density 
can be written as
\beq
     \cL_{\rm SME} = \frac {\sqrt{-g}}{2\ka} 
     (R -2\La+ (k_R)_\abgd R^\abgd )+ \cL^\prime.
\label{lag1}
\eeq
In this expression, 
$R^\abgd$ is the Riemann curvature tensor, 
$\Lambda$ is the cosmological constant, 
$(k_R)_\abgd$ are the SME coefficient fields \cite{k04}, 
and $\ka=8\pi G_N$, where $G_N$ is the gravitational constant.
A generic Lagrangian $\cL^\prime$ appears in the case 
when the coefficients arise dynamically, 
as in spontaneous Lorentz-symmetry breaking. 
The coefficients $(k_R)_\abgd$  can be written
as a scalar $u$, 
two tensor $s_\mn$, 
and four tensor $t_\abgd$ through a Ricci decomposition. 
So we can write
\beq
(k_R)_\abgd R^\abgd = -u R + s_\mn \left(R^{(T)}\right)^\mn + t_\abgd W^\abgd,
\label{ust}
\eeq
where $\left(R^{(T)}\right)^\mn$ is the trace-free Ricci tensor, 
and $W^\abgd$ is the Weyl curvature tensor. 
The coefficients $u$ and $s_\mn$ can be moved to the matter sector
by field re-definitions at first order in these coefficients \cite{bonder15},
so it is important to consider also the matter sector when calculating
physically measureable results.
In this work we choose our conventions such that these coefficients
reside in the gravity sector.

The action obtained from integrating \rf{lag1} over $\int d^4x$
has several features that play a key role in the analysis to follow.
The relevant symmetries in spacetime 
are the four-dimensional diffeomorphism symmetry group 
and local Lorentz transformations. 
Firstly, 
the action is invariant under general coordinate transformations, 
i.e., it is {\it observer} diffeomorphism invariant.
This means specifically that the curvature tensors 
and the coefficients $(k_R)_\abgd$ transform under observer
transformations (change of coordinates) as tensors; 
however, for \emph{particle} 
diffeomorphisms the curvature, 
metric and other fields transform as tensors 
but the coefficients $(k_R)_\abgd$ remain fixed. 
Therefore, the action associated with \rf{lag1} breaks particle diffeomorphisms.

In the standard vierbein formalism where the metric 
is reduced to Minkowski form $\et_{ab}$ at each point 
with a set of four vectors $e^\mu_{\pt{\mu}a}$ via
$\et_{ab} = e^\mu_{\pt{\mu}a} e^\nu_{\pt{\nu}b} g_\mn$, 
similar considerations apply for local Lorentz transformations $\La^a_{\pt{a}b}$.
Therefore, 
we distinguish {\it observer} local Lorentz transformations, 
under which local tensors $R_{abcd}$ transform as well as the coefficients $(k_R)_{abcd}$.
In contrast, 
under {\it particle} local Lorentz transformations, 
the coefficients remain fixed, 
therefore the action also breaks local Lorentz transformations.
Note that the details of these transformations can involve the notion 
of a background vierbein, and so care is required, 
as discussed in Ref.\ \cite{bbw19}.  

The action formed with \rf{lag1} can be interpreted
as explicit symmetry breaking, 
where the coefficients are nondynamical,
or through spontaneous Lorentz symmetry breaking.
In the latter case, 
the underlying model retains the {\it particle} local Lorentz
and diffeomorphism symmetries 
because the coefficients are dynamical fields. 
There must then be a dynamical mechanism, 
such as a potential function of the fields, 
that triggers a vacuum expectation value $\langle (k_R)_{abcd} \rangle$ 
of the fields \cite{ks89bb}.
Upon specifying the vacuum one can still obtain an effective model of the form \rf{lag1}.
Indeed, 
this has been demonstrated for a variety of models with vector 
and tensor couplings to curvature. 
These results have been discussed extensively elsewhere in the literature, 
particularly for spontaneous symmetry breaking \cite{bk05,bkx08}. 

For either Lorentz and diffeomorphism symmetry breaking scenario, 
there are conservation equations which hold based on the action formed from \rf{lag1}.
The field equations for the metric $g_\mn$ obtained from the action take the form
\beq
G^\mn = (T_{\rm ust})^\mn + \ka (T_M)^\mn,
\label{SMEfe}
\eeq
where the explicit form of $(T_{\rm ust})^\mn$ can be found in Ref.\ \cite{k04}, 
and $(T_M)^\mn$ is the energy-momentum tensor obtained from 
the matter sector.
As a consequence of the traced Bianchi identities $\nabla_\mu G^\mn =0$, 
four conservation laws which must hold are given by
\beq
\nabla_\mu (T_{\rm ust})^\mn =- \ka \nabla_\mu (T_M)^\mn.
\label{SMEcl}
\eeq
That these conservation laws hold will be a key point in this work.
There are also $6$ conservation laws associated with local Lorentz symmetry breaking, 
which we do not display here for brevity.
In generality, 
the recent works of Bluhm and collaborators clarify
the differences between explicit and spontaneous local Lorentz
and diffeomorphism symmetry breaking \cite{bluhm15-17,bbw19},
and the intricacies of the conservation laws.
Note that alternatives to Riemann geometry exist, 
such as Riemann-Finsler geometry which 
has recently garnered attention as an additional avenue of pursuit in Lorentz violation theory and phenomenology \cite{finsler}.

\section{3+1 Variables and Decomposition}
\label{sec:ADM}

\subsection{3+1 Basics}
\label{background}

\begin{figure}[h]
\vspace{2mm}
    \includegraphics[width=0.45\textwidth]{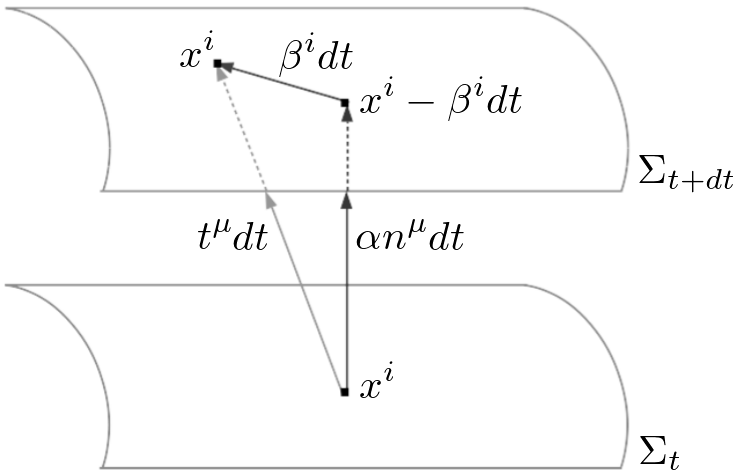}
    \caption{The ADM variables connecting spatial hypersurfaces $\Sigma$ at time $t$ and $t+dt$.}
    \label{fig:ADMvars}
\end{figure}

We start with a 4-dimensional manifold $\mathcal{M}$ with associated metric $g_{\mu\nu}$. 
Following standard methods \cite{adm, deWitt, mtw, boj}, 
we decompose $\mathcal{M}$ into constant-time 
spatial hypersurfaces $\Sigma_t$ with associated timelike normal vector $n^\mu$ 
(normalized to $n_\mu n^\mu = -1$).
In a commonly used coordinate representation the components are $n_\mu = (-\al,0,0,0)$, 
where $\al$ is the Arnowitt-Deser-Misner (ADM) lapse function.
Referring to Figure~\ref{fig:ADMvars}, 
$t^\mu dt$ is a vector connecting points on neighboring hypersurfaces with fixed spatial coordinates $x^i$ (c.f. Figure 2.4 in \cite{numGR}).
The shift vector is $\be^i$, 
which arises from the difference between $t^\mu$ and $n^\mu$, 
while the inverse spatial metric $\gamma^{\mu\nu}$ is given by
\beq
\ga^\mn = g^\mn + n^\mu n^\nu.
\label{gamma}
\eeq
Using $n^\mu$ and $\ga^\mn$, 
the four dimensional curvature $R_\abgd$ is decomposed into 
a three dimensional spatial curvature $\cR_\abgd$ and extrinsic curvature $K_\mn$. 
The extrinsic curvature is defined in terms of the Lie derivative along
$n^\mu$ as
\beq
K_\mn = - \frac 12 \lie \ga_\mn.
\label{ext}
\eeq

A spatial covariant derivative $\mathcal{D}_\mu$ 
is obtained from $\ga$-projection of the covariant derivative of a tensor.
For a tensor with mixed indices $T^\mu_{\pt{\mu}\nu}$, 
for example, 
it is given by
\beq
\cD_\mu T^\al_{\pt{\al}\be} = \ga^\de_{\pt{\de}\mu} \ga^\al_{\pt{\al}\ep} 
\ga^\ze_{\pt{\ze}\be} \nabla_\de T^\ep_{\pt{\ep}\ze}.
\label{3deriv}
\eeq
It will be useful also to use the ``acceleration" $a_\mu=n^\nu \nabla_\nu n_\mu$, 
which is orthogonal to $n^\mu$.
The three-dimensional curvature $\cR^\al_{\pt{\al}\be\ga\de}$ is 
defined by the commutator of the spatial covariant derivatives as
\beq
\left[ \cD_\al, \cD_\be \right] v_\de  = 
-\cR^\ep_{\pt{\ep}\de\al\be} v_\ep,  
\label{3curv}
\eeq
and satisfies $\cR^\ep_{\pt{\ep}\de\al\be} n_\ep =0$, 
where $v_\ep$ is assumed spatial ($n^\ep v_\ep =0$).
Some more key results in the 3+1 decomposition are included 
in the Appendix \ref{sec:app3+1}.

When necessary we will refer specifically to the spacetime metric 
in standard ADM form expressed as a line element,
\beq
ds^2 = -(\al^2 -\be^j \be_j) dt^2 
+2 \be_j dt dx^j +\ga_{ij} dx^i dx^j.
\label{ADMmetric}
\eeq
This form makes plain that the principle variables for gravity are the 
$10$ degrees of freedom $\al$, $\be^j$, and $\ga_{ij}$.

\subsection{GR and the SME action} 
\label{GR and the SME action}

\subsubsection{GR Action}

Of principle importance in what follows 
is that in the SME action, 
spacetime covariant derivatives occur which act on the coefficients for Lorentz violation,
and do not generally vanish.
To see this, 
we decompose the Lagrangian \rf{lag1} using the 3+1 curvature projections 
in the Appendix \rf{curvproj2}.
For reference, 
we examine first the GR Lagrange density which is
\beq
   \cL_{GR} = \frac {\sqrt{-g}}{2\ka}[ \cR + K^\ab K_\ab - K^2 
   - 2 \nabla_\al (n^\al K + a^\al )].
\label{gr3+1}
\eeq
Note here that the last terms form a three-dimensional surface term in the action 
that normally does not affect the dynamical field equations, 
and thus they are usually dropped.\footnote{See Ref.\ \cite{boj} for details
on surface terms.}
What is left contains the extrinsic curvature $K_\ab$, 
which can be seen from \rf{ext} to have time derivatives 
of $\ga_\mn$ via the Lie derivative along $n^\mu$.
Specifically, 
if one evaluates the Lie derivative in \rf{ext} 
one obtains the standard result
\beq
K_{ij} = -\frac {1}{2\al} 
(\prt_t \ga_{\ij} - \cD_i \be_j - \cD_j \be_i ),
\label{Kij}
\eeq
and the other components $K_{0\mu}$ contain no new time derivatives other than those in \rf{Kij}. 
The spatial curvature term in \rf{gr3+1} contains no such time derivatives, 
depending only on the curvature in each spatial hypersurface.

The presence of the time derivatives determine the dynamical variables 
used for a Hamiltonian formulation; in GR, 
only time derivatives of $\ga_{ij}$ occur, 
and thus these six components are the only dynamical
degrees of freedom in the Hamiltonian formulation.\footnote{Note that we adopt 
the first order derivative 
form for the action as much as possible, 
particularly for time derivatives. 
Otherwise, the Hamiltonian formalism gets modified 
to accommodate higher derivatives \cite{genH}. 
This would also occur for models with higher than second derivatives present.}
The other metric degrees of freedom $\al$ and $\be^j$ are nondynamical, 
corresponding to the $4$ gauge degrees of freedom in diffeomorphism symmetry.
This leads to the $4$ primary constraints 
in a Hamiltonian analysis of GR.
Note also that, 
while it does not occur in \rf{Kij}, 
the acceleration $a_\mu$ has only spatial derivatives, 
as it can be shown that 
$a_j  = \prt_j \ln \al$ and $a_0=\be^j a_j$. 

\subsubsection{SME action and global background coefficients}
\label{global}

We next examine the contribution of the $(k_R)_\abgd$ 
coefficients in the SME action. 
Using the general curvature expression 
in the Appendix \rf{curvproj2} 
this term can be manipulated into the form
\bea
    \cL_{k_R} &=& \frac {\sqrt{-g}}{2\kappa} \Big\{ (k_R)_\abgd  
    \Big[ \cR^\abgd -6 K^{\al\ga} K^{\be\de} 
    \nonumber\\
    &&
    \pt{\sqrt{-g}} +4 n^\al n^\ga ( K^{\be\ep} K^\de_{\pt{\de}\ep} 
    - K^{\be\de} K^\ep_{\pt{\ep}\ep})
    +4 a^\be n^\ga K^{\al\de} \Big]
    \nonumber\\
    &&
    \pt{\sqrt{-g}}-4 \Big( n^\al n^\ga (n^\ep K^{\be\de} 
    + \ga^{\de\ep} a^\be ) 
    \nonumber\\
    &&
    \pt{\sqrt{-g}} \pt{-4-} -2 n^\al \ga^{\ga\ep} K^{\be\de}  \Big) 
    \nabla_\ep (k_R)_\abgd
    \Big\}.
\label{klag}
\eea
We can see that a term with the covariant derivative of the coefficients occurs 
while the remaining terms are expressible in terms of the 
extrinsic curvature $K_{ij}$ or the acceleration $a^\mu$.
In general spacetimes this term $\nabla_\ep (k_R)_\abgd$ cannot be made to vanish \cite{k04}.
Since we are interested in the dynamical content of the framework 
we can use the 3+1 decomposition to interpret such terms.

Consider first the simpler case of the covariant derivative 
of a covariant vector $b_\mu$.
Using projection and the definition \rf{3deriv}, 
as well as properties of the Lie derivative along $n^\mu$, 
we can write this in terms of the spatial covariant derivative, 
the Lie derivative, and the extrinsic curvature, 
as
\bea
\nabla_\mu b_\nu &=& \cD_\mu b_\nu - n_\nu \cD_\mu (n^\la b_\la)
-2n_{(\mu} K_{\nu )}^{\pt{\nu}\la} b_\la 
\nonumber\\
&&
- n_\mu n_\nu (a^\la b_\la )
+n_\mu n_\nu \lie (n^\la b_\la) - n_\mu \ga^\be_{\pt{\be}\nu} \lie b_\be. 
\nonumber\\
\label{covb}
\eea
It can be checked using \rf{Dv} that the spatial covariant derivative of $b_\nu$ 
will only contain spatial partial derivatives $\prt_j$ of components of $b_\nu$, 
the functions $\al, \be^j$, the extrinsic curvature $K_{ij}$ 
or the three-dimensional connection coefficients $\3g{i}{j}{k}$, 
the latter of which contain only spatial derivatives of $\ga_{ij}$.
Thus $\cD_\mu$ acting in \rf{covb} cannot introduce any time derivatives 
of the metric functions $\al$ and $\be^j$.
From a geometrical perspective, 
this is because the $\cD_\mu$ derivative describes changes 
in the 3 dimensional hypersurface $\Si_t$. 

Since the acceleration $a_\mu$ depends on spatial derivatives of $\al$, 
we are left with the final two terms 
in \rf{covb} as places where time derivatives of 
$\al$ and $\be_j$ might reside.
The projection of $n^\la b_\la$ can be written as
\beq
n^\la b_\la = \frac {1}{\al} (b_0 - \be^j b_j ), 
\label{ndotb}
\eeq
while its Lie derivative is
\beq
\lie (n^\la b_\la) = 
- \frac {\dot{\al}(n^\la b_\la)+ b_i \dot{\be^i}}{\al^2}  
+ \frac {1}{\al} n^\mu \dot{b}_\mu 
-\frac {1}{\al} \be^j \cD_j (n^\la b_\la).
\label{lie:ndotb}
\eeq
Note the appearance of $\dot{\al}=\prt_t \al$ and 
$\dot{\be^j}=\prt_t \be^j$ for the lapse and shift functions.
This implies that in the Hamiltonian analysis we will generally not obtain 
the usual four primary momentum constraints as in GR.
The final Lie derivative term in \rf{covb} is proportional to 
$\ga^i_{\pt{i}\nu} \lie b_i$, 
which can be shown not to contain time derivatives of the gravitational 
variables $\al$, $\be^j$, and $\ga_{ij}$.

One might immediately suspect that the appearance of 
$\dot{\al}$ and $\dot{\be^j}$ is merely 
a coordinate artifact and can be removed by general coordinate transformations. Indeed, 
the SME maintains general coordinate invariance of the action.
Under a general coordinate transformation, 
$b_\mu$ transforms as a covariant vector:
\beq
b^{\prime}_\mu = \frac {\prt x^\nu}{\prt x^{\prime\mu}} b_\nu,
\label{coord}
\eeq
with other quantities transforming as usual. 
However, 
the quantity $n^\mu b_\mu$ occurring in \rf{lie:ndotb}
is a scalar and is the projection of the hypersurface normal $n^\mu$
along the fixed, 
a priori unknown, background $b_\mu$. 
The appearance of $\al$ and $\be^j$ in \rf{ndotb}, 
in a certain sense,
describes the unknown orientation of the background
and the orientation of the hypersurface geometry, 
the latter being tied to the source of the gravitational fields.
If an alternative coordinate system $x^{\mu\prime}$ 
is chosen so that $n^{\mu\prime} b_{\mu\prime} = b_{0'}$
and we then suppose that in the new coordinate system
$b_{\mu\prime}$ is now {\it the} fixed background 
that is independent of the gravitational fields, 
we have effectively made a different choice of background, 
and because of the explicit breaking of diffeomorphism symmetry, 
we have chosen a different model.\footnote{This choice corresponds to Gaussian normal coordinates \cite{mtw}.}
We return to this point later below
when we consider alternative ways of specifying the background fields.

The vector example can be extended to general tensors; 
since our focus is on the SME gravity action, 
in the minimal case, 
it is possible to manipulate the Lagrange density into a form 
where time derivatives dependencies are more transparent 
just like \rf{covb}. 
In fact, we can write \rf{klag} as
\bea
    \cL_{k_R} &=& \frac {\sqrt{-g}}{2\ka} 
    \Big\{ (k_R)_\abgd  \Big[ \cR^\abgd + 2 K^{\al\ga} K^{\be\de} 
    \nonumber\\
    &&
    -12 n^\al n^\ga K^{\be\ep} K^\de_{\pt{\de}\ep} 
    +4 n^\al n^\ga K^{\be\de} K^\ep_{\pt{\ep}\ep} 
    +8 K^{\al\ga} n^\be a^\de \Big]
    \nonumber\\
    &&
    \pt{\sqrt{-g}} 
    +8 K^{\ep\ze} \cD_\la \big( (k_R)_\abgd \ga^\al_{\pt{\al}\ep} \ga^{\be\la} \ga^\ga_{\pt{\al}\ze} n^\de \big) 
    \nonumber\\
    &&
    \pt{\sqrt{-g}} 
    -4 a^\ep \cD_\ze \big( (k_R)_\abgd \ga^{\al\ze} \ga^\ga_{\pt{\al}\ep} n^\be n^\de \big)
    \nonumber\\
    &&
    \pt{\sqrt{-g}}
    -4 K^{\ep\ze} \lie 
    \big( (k_R)_\abgd \ga^\al_{\pt{\al}\ep} 
    \ga^\ga_{\pt{\al}\ze} n^\be n^\de \big) 
    \Big\}.
\label{klag2}
\eea
Any nonstandard time derivative terms will be contained in the last Lie derivative term.
The appearance of $\dot{\al}$ and $\dot{\be^j}$ terms can be verified by working out the Lie derivative term explicitly.
We find the relevant piece to be
\beq
\cL_{k_R} \supset \frac {4 \sqrt{-g}}{\ka \al^2}
K^{ij} n^\de \big((k_R)_{i\be j \de} n^\be \dot{\al} + 
(k_R)_{ilj\de} \dot{\be^l} \big),
\label{lie:dotk}
\eeq
Like in equation \rf{lie:ndotb} we obtain a combination of $\dot{\al}$ and $\dot{\be^j}$ terms, and we thus expect this to hold for general tensors 
in the SME \cite{bkx15}.

This rather interesting result, 
the occurrence of $\dot{\al}$ and $\dot{\be^j}$, 
is in contrast with GR and many modified models of gravity.
It is somewhat unsurprising in that we are considering 
the SME framework interpreted in the context of 
explicit diffeomorphism symmetry breaking, which breaks the gauge symmetry of GR.
Other models, 
such as massive gravity, 
which also have explicit diffeomorphism breaking, 
modify mass-type terms with no derivatives in them. 
They generally do not modify the kinetic structure of the theory 
and thus do not introduce such terms.
As another example, 
for models with curvature contractions in the Lagrangian like $R_\ab R^\ab$, 
even though they have higher than second derivatives of the metric, 
the lapse and shift functions remain gauge~\cite{R2models}.
More varied results exist for other models with higher than second derivatives
\cite{pertAlBe}.

In the case of spontaneous symmetry breaking, 
where for example $b_\mu \rightarrow B_\mu$ is dynamical, 
there are separate field equations for $B_\mu$ 
which must also be considered.  
The net effect in this case, 
since the underlying diffeomorphism symmetry remains,
is that $\dot{\al}$ and $\dot{\be^j}$ can be eliminated
by a particle diffeomorphism.
Or alternatively, 
one can see that the dynamics of $\al$ and $\be^j$ become 
linked to the field $B_\mu$, 
as the time derivatives always occur in the combination in \rf{lie:ndotb},
and they thus do not represent independent degrees of freedom.
This point about the spontaneous-breaking case parallels 
the reasoning behind the observation that 
any diffeomorphism Nambu-Goldstone modes $\Xi^\mu$ vanish 
from terms in the action like $\nabla_\mu B_\nu$ \cite{bk05,bkx08}.

\subsubsection{Local background coefficients}

In the explicit breaking case considered above, 
the coefficients $k_\abgd$ with spacetime indices
are considered as the fixed background fields
independent of the gravitational variables.
There are alternative choices that could change the results. 
For instance, 
using the vierbein formalism it may be more natural 
to treat the coefficients in a different way.
From the basic definition of the vierbein, 
we can find its 3+1 components from the metric \rf{ADMmetric}.
The components $e_\mu^{\pt{\mu}a}$ are given by
\bea
e_t^{\pt{t}\bar{0}} &=& \al,
\nonumber\\
e_j^{\pt{j}\bar{0}} &=& 0,
\nonumber\\
e_t^{\pt{t}\bar{j}} &=& e_i^{\pt{i}\bar{j}} \be^i,
\nonumber\\
\ga_{ij} &=& e_i^{\pt{i}\bar{j}} e_{j \bar{j} } 
\label{ADMvierbein},
\eea
where we use $t$ and $i,j, ...$ for time and space indices while 
for the local frame we use a bar over the index.
The last equation merely defines the spatial piece of the vierbein $e_i^{\pt{i}\bar{j}}$, 
since we have not specified the spatial metric.  
The explicit decomposition can be performed once a spatial metric is chosen.
The vierbein in \rf{ADMvierbein} is not unique; 
one may apply a local Lorentz transformation $\La^a_{\pt{a}b}(x)$ 
and generally mix components.

Returning to the vector example above, 
when using the vierbein it is natural to consider 
the local covariant vector field $b_a$ 
as the fundamental 
background object which breaks
the spacetime symmetries \cite{k04,c19}.
For instance, using the vierbein and the vector $n^\mu$ the projection which occurs in equation \rf{lie:ndotb} can be written
\beq
b_\mu n^\mu = b_{\bar{0}}.
\label{blocal}
\eeq
In this case the Lie derivative term in \rf{lie:ndotb} yields
\beq
\lie (n^\la b_\la) = 
\frac {\prt_t b_{\bar{0}}}{\al}
-\frac {1}{\al} \be^j \prt_j (b_{\bar{0}}).
\label{lie:ndotb2}
\eeq
Now we can see that no time derivatives of $\al$ and $\be^j$ occur, 
provided that $b_a$ is {\it the} independent background.
\footnote{An alternative way to arrive at equation \rf{blocal} 
is to define ${\bf n} \cdot {\bf b}= b_\mu n^\mu = b_{\perp}$ 
as {\it the} time component \cite{IN77}.}
It should be noted that this choice does
not make use of a background vierbein, 
as discussed in Ref.\ \cite{bbw19}, 
and may result in more severe constraints
on explicit breaking models via the conservation laws.

How does all of this play into the dynamical and propagation structure that is known from weak-field studies of the SME
and models of spacetime symmetry breaking? 
To answer this we also perform a comparison 
with what is known about the
weak field quadratic limit \cite{km18}, 
including generic gauge-violating terms
in Section \ref{sec:quadSME}.
Ultimately in this work, 
we look at cases of explicit breaking with 
both choices of the background coefficients corresponding
to the ``global" background in \rf{ndotb} and the 
``local" background \rf{blocal}.

\section{Hamiltonian Analysis}
\label{sec:ham}

\subsection{Generalities}
\label{generalities}

Working with lagrange density in the form of Eq.~\rf{klag2}, 
we carry out a Riemann decomposition of $(k_R)_\abgd$
into $u$, $s_\mn$, and $t_{\ka\la\mu\nu}$. 
The Lagrange density for $t_{\ka\la\mu\nu}$ 
will take the same form as $\rf{klag2}$ with the replacement of $(k_R)_\abgd \rightarrow t_\abgd$.
For $u$ and $s_\mn$ we obtain
\bea
\cL_u &=& \frac {\sqrt{-g}}{2\ka} [ u \left( \cR + K^\ab K_\ab - K^2 \right) 
\nonumber\\
&&
\pt{space}+2 ( K \lie u + a^\mu \cD_\mu u ) ]
\nonumber\\
\cL_s &=& \frac {\sqrt{-g}}{2\kappa} [ s_\mn \cR^\mn  
- n^\al n^\be s_\ab (K^\mn K_\mn - K^2 )
\nonumber\\
&& \pt{-g} 
+2 s_\ab K^{\al\de} K^\be_{\pt{\be}\de} 
+K^\mn \lie s_\mn 
-K \lie (n^\mu n^\nu s_\mn) 
\nonumber\\
&& \pt{-g} 
+ 2 K \left( s_\mn n^\mu a^\nu + \cD_\la (s_\mn n^\mu \ga^{\nu\la})\right) 
\nonumber\\
&&\pt{-g} 
- 2 K^\la_{\pt{\la}\ka} \cD_\la (s_\mn n^\mu \ga^{\nu\ka} )
\nonumber\\
&&\pt{-g} 
+ a_\ka \cD_\la (s_\mn \ga^{\mu\la} \ga^{\nu\ka})
- a^\la \cD_\la ( s_\mn n^\mu n^\nu )].
\label{us}
\eea
Note that one can also consider other possibilities 
like substituting $u \rightarrow g^\mn s_\mn$.

In the Hamiltonian analysis, 
one first finds the canonical momentum densities using $L=\int d^3x \cL$ via the standard variational definition
\beq
\Pi_n = \frac {\de L}{\de \dot{\ph}_n}.
\label{pdef}
\eeq
In the present case the $\ph_n$ correspond to $\al$, $\be^i$, and $\ga_{ij}$.
To describe the results for the SME actions 
we show the canonical momenta for the $u$, $s_\mn$, and $t_\abgd$ terms, 
assuming global background coefficients as in Sec.\ \ref{global}.

For $\Pi_\al = \de L/\de \dot{\al}$ and $(\Pi_\be)_i = \de L/\de \dot{\be}^i$ we obtain
\bea
\Pi_\al &=& \frac{\sqrt{\ga}}{ \ka \al} n^\mu n^\nu \left( 
K s_\mn + 4 K^{ij} t_{i\mu j \nu} \right),
\\
\Pi_{\be,i} &=& \frac{\sqrt{\ga}}{ \ka \al} n^\mu \left( 
K s_{\mu i} + 4 K^{jk} t_{jik\mu} \right).
\label{Piab}
\eea
Note that no nonzero terms appear here for the case of $u$; 
however, if one chooses $u$ to be composite, such as $u=s^\mu_{\pt{\mu}\mu}$, a different result ensues.
For the $s_\mn$ and $t_{\ka\la\mu\nu}$ coefficients the expressions for $\Pi^{ij}_\ga=\de L /\de \dot{\ga}_{ij}$
are lengthy and omitted here.
These expressions contain terms which generally mix the components of $s_{ij}$ and $\dot{\ga}_{ij}$
in an anisotropic manner; 
for instance, 
\beq
\Pi^{ij}_\ga \supset \tfrac{\sqrt{\ga}}{ 2 \ka} 
\left( K \ga^{ij} - K^{ij} - s_{lm} \ga^{li} K^{jm} 
- s_{lm} \ga^{lj} K^{im} + ... \right),\nonumber\\
\label{Pisample}
\eeq
where the first two terms are the $GR$ result and the displayed remaining terms 
show a mixing of the components of $s_{ij}$ and $K_{ij} \sim \dot{\ga}_{ij}-\cD_i \be_j - \cD_j \be_i$.

To construct the Hamiltonian density $\cH = \Pi^n \ph_n - \cL$, 
one needs to express the $\dot{\ph}_n$ in terms of
the momenta $\Pi_n$.
Since obtaining the general expression for $\dot{\ga}_{ij}$ 
involves a lengthy process of inversion  
due to the anisotropic components of the coefficients, 
we endeavour in this work to begin an investigation by 
studying special limiting cases of the underlying action.

\subsection{Case study 1}
\label{case study 1}

We consider a special case with one nonzero component $s_{00}$ of $s_\mn$ 
in the chosen coordinate system.\footnote{Note that 
alternative choices exist such as considering 
the contravariant coefficients as the fixed background; for example, 
$n_\mu n_\nu s^\mn = \al^2 s^{00}$ for arbitrary $s^\mn$.} 
In this case, 
using the specific components of the metric \rf{ADMmetric} 
the Lagrange density simplifies to
\bea
    \cL_1 &=& \frac {\al \sqrt{\ga}}{2\ka} \Big[
    \cR +\frac{\al^2-s_{00}}{\al^2}\left(K^{ij}K_{ij} -K^2 \right)
    \nonumber\\
    &&
    \pt{\al\ga} 
    +K \left( \frac {2}{\al^4} s_{00} (\dot{\al} - \al \be^i a_i )
    -\frac {1}{\al^3} (\dot s_{00} - \be^i \prt_i s_{00}) \right)
    \nonumber\\
    &&
    \pt{\al\ga} 
    + \frac {2}{\al^2} s_{00} a^i a_i - \frac {1}{\al^2} a^i \prt_i s_{00}  
    \Big]+ \cL_M.
\label{case1}
\eea
When constructing the Hamiltonian, 
the variables of the system are the $\alpha$, $\beta^\mu$, 
and $\gamma_{\mu \nu}$ fields along with their conjugate momentum densities:
\bea
   \Pi_\ga^{ij} &=&\frac {\sqrt{\ga}}{2\ka} 
   \Big[\frac{\al^2-s_{00}}{\al^2}
   \left(K\ga^{ij}-K^{ij}\right)
   \nonumber\\
   && \pt{\sqrt{ga}}
    +\frac 1{2\al} \ga^{ij} (\prt_t - \be^k \prt_k) 
    \left( \frac {s_{00}}{\al^2} \right) \Big],
    \label{piG} \\
    \Pi_{\be,i} &=&0,
    \label{PC}\\
    \Pi_\al &=&\frac{\sqrt{\ga}s_{00}}{\ka \al^3}K.
    \label{PA}
\eea
From now on we drop the $\ga$ label on $\Pi^{ij}$ 
and abbreviate its trace as $\Pi=\Pi^{ij} \ga_{ij}$, 
and we drop the $\be$ label so that $\Pi_i=\Pi_{\be,i}$.

Examining the momenta, 
we see that equation \rf{PC} gives three primary constraints.
The equations \rf{piG} and \rf{PA}, 
along with equation \rf{Kij}, 
can be inverted to solve for $\dot{\ga}_{ij}$ and $\dot{\al}$.
Following standard procedure \cite{IN77}, 
we first find the {\it base} Hamiltonian density of the system 
through a Legendre transformation on the Lagrange
density $\cH_0=\pi^{ij} \dot{\ga}_{ij}+\pi_{\al}\dot{\al}-\cL$. 
For the base Hamiltonian density we find
\bea
    \cH_0 &=& \frac{2\ka \al^3}{\sqrt{\ga}(\al^2- s_{00})} 
    \left(\Pi_{ij}\Pi^{ij}
    -\frac{1}{3}\Pi^2 \right)
    \nonumber\\
    &&
    +\frac{ \ka \al^5 (\al^2 - s_{00})}{3\sqrt{\ga}s_{00}^2}\Pi_\al^2
    -\frac{2 \ka \al^4 }{3 \sqrt{\ga} s_{00}} \Pi_\al \Pi 
    +\frac { \al \dot{s}_{00} }{2 s_{00}} \Pi_\al
    \nonumber\\
    &&
     -\frac {\sqrt{\ga} }{\ka \al} 
     \left( s_{00} a^i a_i - \frac 12 a^i \prt_i s_{00} \right)
    -\frac {\al \sqrt{\ga}}{2 \ka} \cR
    \nonumber\\
    &&
    +\be^i \left( \Pi_\al [\al a_i - \frac {\al}{2 s_{00}} \prt_i s_{00}] 
    -2 \cD_j \Pi^j_{\pt{j}i} \right).
    \label{H0}
\eea
Note that by comparison, in GR, the momentum for $\al$ is absent, thus Eq.\ \rf{PA} is replaced with $\Pi_\al=0$.
The GR Hamiltonian is
\beq
\cH_{\rm GR}= \fr {2\ka \al}{\sqrt{\ga}} 
\left(\Pi_{ij}\Pi^{ij}
    -\frac{1}{2}\Pi^2 \right) - \fr {\al \sqrt{\ga}}{2\ka} \cR
    -2 \be^i \cD_j \Pi^j_{\pt{j}i}.
\label{GRham}    
\eeq
Examining \rf{GRham} and \rf{H0} it can be seen that taking the
limit of $s_{00} \rightarrow 0$ does not smoothly connect the 
Hamiltonians.  
This is an artifact of the Hamiltonian method, 
in particular it results from solving Eq.\ \rf{PA} for $\ga^{ij} \dot{\ga}_{ij}$, 
which requires $s_{00} \neq 0$.  
We return to this point later in the process below.

To Eq.\ \rf{H0} we add a term involving 
the primary constraint contracted 
with a Lagrange multiplier to obtain the {\it augmented} Hamiltonian 
$\cH_A = \cH_0 + \xi^i \Pi_i$. 
We then check the consistency condition, 
or evolution,
for this primary constraint
by taking its Poisson bracket with the augmented Hamiltonian
$\dot{\Pi}_i = \{ \Pi_i, H_A \}$.
This yields a secondary constraint
\beq
\dot{\Pi}_i=2 \ga_{ij} \cD_k \Pi^{jk} -\Pi_{\al} 
\left( \al a_i  - \frac {\al}{2 s_{00}} 
\prt_i s_{00} \right) \approx 0.
\label{secondary}
\eeq
Note that the $\approx$ symbol here refers to an expression
that is ``weakly" equal to zero, 
i.e., when the constraints are imposed it vanishes \cite{dirac}.
This secondary constraint can also be observed in the last line
of equation \rf{H0} multiplying the $\be^i$.

We continue to check consistency conditions with the secondary constraint 
$\Ph_i=\dot{\Pi}_i$.
The full expression for the evolution of $\Ph_i$ is needed, 
including the explicit time dependence since there may be additional time 
dependence in $s_{00}$.
A lengthy calculation reveals
\bea
\frac {d \Ph_i}{dt} &=& \{ \Ph_i, H_A \} 
+ \frac {\prt \Ph_i}{\prt t} \nonumber\\
&=&\cD_j ( \be^j \Ph_i ) + \Ph_j \cD_i \be^j + \Ps \prt_i s_{00},\nonumber\\
\label{tert}
\eea
where $\Ps$ is a function of the coordinates and momenta equal to
\bea
\Ps &=& -\frac{\ka \al^3}{\sqrt{\ga}(\al^2- s_{00})s_{00}} 
    \left(\Pi_{ij}\Pi^{ij} -\frac{1}{3}\Pi^2 \right)
    \nonumber\\
    &&
    -\frac{ \ka \al^5 (\al^2 - s_{00})}{6\sqrt{\ga}s_{00}^3} \Pi_\al^2
    +\frac{ \ka \al^4 }{3 \sqrt{\ga} s_{00}^2} \Pi_\al \Pi 
    +\frac {\al \sqrt{\ga}}{4 \ka s_{00}} \cR
    \nonumber\\
    &&
    +\frac {\sqrt{\ga}}{4 \ka s_{00} \al} \cD^2 s_{00} 
    -\frac {\sqrt{\ga}}{2 \ka \al^2} \cD^2 \al
    \nonumber\\
    &&
    -\frac {\sqrt{\ga}}{\ka s_{00} \al} a^j \prt_j s_{00}
    +\frac {3 \sqrt{\ga}}{2 \ka \al} a^j a_j.
\label{ps}
\eea
The implications of \rf{tert} are as follows.
Examining this expression the first two terms 
are linear in the original constraint $\Ph_i$, 
and so are weakly equal to zero - providing no new constraints. 
The last terms would appear to give new constraints, 
but this depends on the properties of the background coefficients $s_\mn$.
If we insisted that the coefficients and their derivatives remain arbitrary 
we would have to take the last terms in \rf{tert} 
as new constraints and again check the consistency 
using Poisson brackets with the Hamiltonian.
On the other hand, 
if we merely insist that in the chosen coordinate system 
$\prt_i s_{00} = 0$ 
then \rf{tert} will be weakly equal to zero 
and no new constraints are needed.
That a constraint on $s_{00}$ has arisen directly 
from this analysis can be traced to the fact that \rf{secondary} 
is a modification of the usual momentum constraint of GR. 
The term $\prt_i s_{00}$ represents an additional 
kind of ``shift" in the momentum conservation law.

Further insight can be gained by examining the traced Bianchi
identities \rf{SMEcl} for the choice of coefficients we have made.
From reference \cite{k04} we have 
\beq
\label{eq:bianchivacuum}
     \nabla_\mu (T_s)^\mu_{\pt{\mu}\nu} =
     \frac{1}{2}R^{\mu\la} \nabla_\nu s_{\mu\la}
     -\nabla_\mu (R^{\mu\la} s_{\la\nu}). 
\eeq
We next re-express this equation in 3+1 form by using Appendix Eq.\ \rf{4Ricci}, 
the decomposition of the covariant derivative like in Eq.\ \rf{covb},
and we use equations \rf{piG}-\rf{PA} to write the expression in 
terms of the Hamiltonian variables 
$\al$, $\be^j$, $\ga_{ij}$, $\Pi_\al$, and $\Pi^{ij}$.
Examining the case of $s_{00}$ only then yields
\beq
\nabla_\mu (T_s)^\mu_{\pt{\mu}j} = \frac {\ka}{\sqrt{\ga}\al} \Ps \prt_j s_{00}, 
\label{conslaw1}
\eeq
which contains the same terms as in \rf{tert}.
Therefore, 
we see that the Hamiltonian evolution has produced
a constraint that we expect from the field equations.

We will proceed with the assumption that the coefficients $s_{00}$ 
are independent of spatial coordinates:
\beq
\prt_ i s_{00} = 0.
\label{s00Cond}
\eeq
Note that this is a coordinate-dependent statement 
which may be more properly understood as saying that $s_{00}$
does not change within the spatial hypersurface at fixed $t$.

Hamilton's equations of motion can now be obtained in the standard way through the Poisson bracket 
$\dot{p}_n = \left\{p_n , H \right\}$, 
where $H$ is the {\it final} Hamiltonian with the primary constraint added.
In principle, 
one adds the secondary constraint to the Hamiltonian
with an additional three Lagrange multipliers.
Since the secondary constraint can be seen to be already contained in the $\be^j$ term
in the Hamiltonian \rf{H0}, 
it is not strictly necessary to add this term.
This reflects the remaining gauge freedom in this limit of the framework.

We then find the Hamilton's equations of motion for the momentum variables to be
\begin{widetext}
\bea
    \dot{\Pi}_\al &=& 
    -\frac{2 \ka \al^2 (\al^2-3s_{00})}{\sqrt{\ga} (\al^2-s_{00})^2}
    \left( \Pi^{ij}\Pi_{ij} -\frac{1}{3}\Pi^2 \right)
    + \frac{8 \ka \al^3}{3\sqrt{\ga}s_{00}} \Pi_\al \Pi 
    - \frac{\ka \al^4}{3\sqrt{\ga}s^2_{00}}
    \left(7\al^2 - 5s_{00} \right)\Pi_\al^2 
    + \cD_k \left(\be^k\Pi_\al \right) 
    \nonumber\\
    && 
    - \frac{1}{2 s_{00}}\Pi_\al \dot{s}_{00} 
    +\frac{s_{00}\sqrt{\ga}}{\ka \al^2} \left( a^i a_i - 2 \cD_i a^i \right)
    +\frac {\sqrt{\ga}}{2 \ka} \cR
    \label{eq:eom_alpha},
    \\
    \dot{\Pi}_i &=& 2 \ga_{ij} \cD_k \Pi^{jk} -\Pi_{\al} \al a_i ,
    \label{eq:eom_beta} \\
    \dot{\Pi}^{ij} &=& 
    -\frac{4\ka \al^3}{\sqrt{\ga}\left(\al^2-s_{00}\right)}
    \left(\Pi^i_{~k}\Pi^{jk} -\frac{1}{3}\Pi \Pi^{ij}\right)
    +\frac{\ka \al^3}{\sqrt{\ga}(\al^2-s_{00})}\ga^{ij}
    \left(\Pi^{kl}\Pi_{kl}  -\frac{1}{3}\Pi^2\right)
    + \frac{\ka \al^5 (\al^2- s_{00})}{6 \sqrt{\ga} s_{00}^2} \ga^{ij} \Pi_\al^2
    \nonumber\\
    &&
    - \frac{\ka \al^4}{3 \sqrt{\ga} s_{00}} \Pi_\al 
    (\Pi \ga^{ij} - 2\Pi^{ij} ) 
    - 2\Pi^{k(i} \cD_k \be^{j)} 
    +\cD_k\left( \Pi^{ij}\be^k \right)
    -\frac{\sqrt{\ga}}{2\ka} 
    \left( \al \cR^{ij}-\frac{1}{2}\ga^{ij}\al \cR 
    -\cD^i \cD^j \al + \ga^{ij} \cD^2 \al \right)
    \nonumber\\
    &&
    + \frac{ \sqrt{\ga} s_{00}}{\ka \al} 
    \left( \frac 12 \ga^{ij} a^k a_k - a^i a^j \right), 
    \label{eq:eom_gamma}
\eea
\end{widetext}
and
\bea
    \dot{\al} &=& 
    -\frac{2\ka \al^4}{3s_{00}\sqrt{\ga} }
    \left[ \Pi - \frac{\al(\al^2-s_{00})}{s_{00}} \Pi_\al \right]
    +\al \be^k a_k 
    \nonumber\\
    &&
    + \frac {\al}{2 s_{00}} \dot{s}_{00} 
    \\
    \dot{\be}^i &=& \xi^i \\
    \dot{\ga}_{ij} &=& \frac{4 \ka \alpha^3}{\sqrt{\ga}(\al^2-s_{00})}
    \left(\Pi_{ij} -\frac{1}{3} \Pi \ga_{ij}\right)
    \nonumber\\
    &&
    -\frac{2\ka \al^4}{3\sqrt{\ga} s_{00} } \Pi_\al\ga_{ij}
    +\cD_i \be_j+\cD_j \be_i.
    \label{eom:qs}
\eea
Note that we have implemented the condition \rf{s00Cond}.

At this point it is useful to remark upon the degrees of freedom 
in this special case of the SME.\footnote{
Degrees of freedom represent
pairs of coordinate and momenta variables freely specifiable
on a hypersurface of fixed $t$ \cite{ms1}.}
We began with up to $10$ degrees of freedom in the variables $\al$, 
$\be^j$, 
and $\ga_{ij}$.
With our choice of $s_{00}$ we have three primary constraints 
and three secondary constraints, 
along with six undetermined Lagrange multipliers.
According to the standard recipe 
(see for example, equation B11 in Appendix B of Ref.\ \cite{IN77}) 
one can use the equation 
\bea
N_{\rm dof} &=& N_{\rm dof, initial} - \tfrac 12 ({\rm \# constraints} )
\nonumber\\
&&
-\tfrac 12 ({\rm \# undetermined \, Lagrange \, multipliers} ),
\label{dof}
\eea
to determine the number of degrees of freedom.
In the case above, 
we have $10 - (1/2)(6) - (1/2)(6) = 4$ degrees of freedom in our model.
In GR, by contrast, there are $4$ primary constraints, 
$4$ secondary constraints, 
and in principle $8$ undetermined Lagrange multipliers
which leaves $10- (1/2)8 - (1/2)8 =2$ degrees of freedom. 

Note the appearance of the inverse of $s_{00}$
in the expressions above.
This does not represent a phase space singularity, 
but rather a parameter singularity.
Its appearance is tied to the Hamiltonian method, 
where one inverts, 
for example, 
equation $\rf{PA}$,
while in contrast results are generally linear 
in the parameter $s_{00}$ in the standard Euler-Lagrange equations.
There is also a denominator in \rf{H0} that appears to have a singular point when $\al=\pm \sqrt{s_{00}}$.
It is not clear if anything keeps the field configurations
away from this region of phase space and this remains to be investigated.

The Hamiltonian and Hamilton's equations in this special example. 
or a related one with different choices of $s_{\mu\nu}$,
can form the basis for future work in a variety of areas such as
studying the initial value formulation of the system of equations.
This could lead to modeling the effects of SME coefficients 
in strong field gravity systems, 
for example using numerical techniques of integration \cite{numGR}.
In this paper, 
we content ourselves with a cosmology application in Section \ref{sec:cos}.

\subsection{Case study 2}
\label{case study 2}

In contrast the case considered above, 
we can make an alternative choice for the background coefficients. 
In this example we choose $s_{ab}$ to be diagonal and isotropic 
in the local Lorentz frame 
\bea
s_{ab} = \left( 
\begin{array}{cccc} 
\sl & 0 & 0 &0 \\
0 & \fr 13 s & 0 & 0\\
0 & 0 & \fr 13 s & 0 \\
0 & 0 & 0 & \fr 13 s \\
\end{array}
\right),
\label{local}
\eea
where the nonzero components $\sl$ and $s$ 
are left as arbitrary functions of the spacetime.
Using the vierbein in \rf{ADMvierbein} we can find 
the components $s_\mn=e_\mu^{\pt{\mu}a} e_\nu^{\pt{\mu}b} s_{ab}$ 
in the spacetime coordinates of the metric \rf{ADMmetric}.
Simplifying the action for $s_\mn$ in \rf{us} we obtain an 
alternative explicit breaking Lagrangian:
\bea
    \cL_2 &=& \frac {\al \sqrt{\ga}}{2\ka} 
    \Big[
    \cR \left(1+ \tfrac 13 s \right) + \left(K^{ij}K_{ij} -K^2 \right)(1-\sl)
    \nonumber\\
    &&
    \pt{\al\ga} 
    +K \lie \Om 
    + a^i \prt_i \Om \Big],
\label{case2}
\eea
where we use the abbreviation $\Om = s/3 - \sl$.

In this case the terms involving the time derivatives of $\al$ and $\be^j$ are absent, 
and except for the time and space dependence of the coefficients which we take as 
arbitrary for the moment, 
the Lagrange density resembles that of GR with scalings of the extrinsic curvature 
and spatial curvature terms.
The canonical momenta are calculated to be
\bea
   \Pi^{ij} &=&\frac {\sqrt{\ga}}{2\ka} 
   \Big[\left(K\ga^{ij}-K^{ij}\right)(1-\sl )
     - \tfrac 12 \ga^{ij} \lie \Om \Big],
    \label{piG2} \\
    \Pi_{\be,i} &=&0,
    \label{PC2}\\
    \Pi_\al &=&0.
    \label{PA2}
\eea
Note the appearance of the Lie derivative of the coefficients directly 
in the momentum and that we get four primary constraints for $\al$ and $\be^j$, 
as in GR.
This is in contrast to the case of global coefficients in the general analysis of section \ref{generalities}, 
and so \rf{PA2} is not a limit of Eq.\ \rf{Piab}. 
The base Hamiltonian density for this case is given by
\bea
\cH_0 &=& \frac{2\ka \al}{\sqrt{\ga}(1-\sl)} 
    \left(\Pi_{ij}\Pi^{ij} -\frac{1}{2}\Pi^2 \right)
    +2 \Pi^{ij} \cD_i \be_j 
    \nonumber\\
    &&
    - \frac {\al \sqrt{\ga}}{2 \ka} \left(1+\tfrac 13 s\right) \cR
    -\fr {1}{2 (1 - \sl )} \Pi \Om^\prime 
    \nonumber\\
    &&
    -\frac {3 \sqrt{\ga}}{16 \ka \al (1-\sl)}
    (\Om^\prime)^2
    -\frac {\al \sqrt{\ga}}{2 \ka} a^i \prt_i \Om
\label{H02}
\eea
where for convenience we define $\Om^\prime  = (\prt_0 - \be^i \prt_i) \Om $.

The evolution of the primary constraints 
with respect to the augmented Hamiltonian
\beq
H_A = \int d^3 x (\cH_0 + v \Pi_\al + \xi^i \Pi_i),
\label{HA2}
\eeq 
where $v$ and $\xi^j$ are lagrange multipliers,
yields the following secondary constraints:
\bea
\{ \Pi_\al , H_A \} &=& -\frac {2 \ka}{\sqrt{\ga} (1 - \sl)} 
\left( \Pi^{ij} \Pi_{ij} - \tfrac 12 \Pi^2 \right) 
\nonumber\\
&&
+\frac {\sqrt{\ga}}{2 \ka} (1+\tfrac 13 s) \cR
- \frac {3 \sqrt{\ga}}{16 \ka (1-\sl) \al^2} (\Om^\prime)^2 
\nonumber\\
&& - \frac {\sqrt{\ga}}{2\ka} \cD^2 \Om
\label{secondary2a},\\
\{ \Pi_i , H_A \} &=& 2 \cD_j \Pi^j_{\pt{j}i} - \frac {\Pi}{2 (1-\sl)} \prt_i \Om
\nonumber\\
&&
-\frac {3 \sqrt{\ga}}{8 \ka (1-\sl) \al} \Om^\prime \prt_i \Om .
\label{secondary2b}
\eea

These secondary constraints contain the GR secondary constraints
but they differ in the extra terms 
involving time and space derivatives of the coefficients in $\Om$.
The standard procedure is to check the consistency of these secondary constraints.
Due to the presence of the spatial derivatives in $\Om$, 
we expect a result similar to that for the case 1 model, 
whereupon we obtain a lengthy function of the canonical variables
multiplied by terms proportional to $\prt_i \Om$. 
This is indeed confirmed by calculation, 
and so we proceed with the simplifying assumption that $\prt_i \Om =0$.
This assumption has the immediate effect of 
reducing the secondary constraints in \rf{secondary2b} 
for $\Pi_i$ to that of the standard ones for GR, 
$\dot{\Pi}_i=2\cD_j \Pi^j_{\pt{j}i} \approx 0$.

Still allowing for arbitrary time dependence of the coefficients $\sl$ and $s$,
we proceed with the calculation of the secondary constraint evolution.
Denoting $\Ph_\al = \{ \Pi_\al , H_A \}$, and $\Ph_i = \{ \Pi_i , H_A \} $, 
we obtain the following results for their evolution:
\begin{widetext}
\bea
\{ \Ph_\al, H_A \} + \frac {\prt \Ph_\al}{\prt t} &=& \cD_i (\be^i \Ph_\al ) 
+ \frac {2 (1+\tfrac 13 s)}{(1-\sl)} \Ph_i \cD^i \al
+ \frac { (1+\tfrac 13 s)\al}{(1-\sl)} \cD^i \Ph_i
+ v \frac {3 \sqrt{\ga} }{8 \ka \al^3 (1-\sl ) } \dot {\Om}^2
+ \frac {9 \sqrt{\ga} }{64 \ka \al^2 (1-\sl )^2 } \dot {\Om}^3
\nonumber\\
&&
+ \frac {3}{8 \al (1-\sl )^2  } \dot {\Om}^2 \Pi 
-\frac { \ka (\dot{s} + \dsl ) } {2\sqrt{\ga} (1-\sl )^2} 
\left(\Pi_{ij}\Pi^{ij} -\frac{1}{2}\Pi^2 \right)
- \frac {3 \sqrt{\ga} }{8 \ka \al^3 (1-\sl ) } \dot {\Om}^2 \be^i \cD_i \al
\nonumber\\
&&
-\frac {\sqrt{\ga}}{8 \ka (1-\sl)} \left((1+ \tfrac 13 s) \dot{\Om}
- \tfrac 43 \dot{s} (1-\sl ) \right) \cR
- \frac {3 \sqrt{\ga} }{8 \ka \al^2 (1-\sl )} \dot {\Om} \ddot{\Om}
- \frac {3 \sqrt{\ga} } {16 \ka \al^2 (1-\sl )^2 } \dot{\Om}^2 \dsl,
\label{Phialdot2}\\
\{ \Ph_j, H_A \} &=& \Ph_i \cD_j \be^i + \cD_i (\be^i \Ph_j ) + \Ph_\al \cD_j \al.
\label{Phiidot2}
\eea
\end{widetext}

Examining these expressions reveals two things.
Firstly, 
from \rf{Phiidot2},
we see that the secondary constraint $\cD_i \Pi^i_{\pt{i}j}$ 
is preserved since its evolution 
is linear in the secondary constraints, 
which weakly vanish.
Note in particular in \rf{Phiidot2} 
that terms linear in $\Ph_i$ {\it and $\Ph_\al$} appear, 
in contrast to case study 1 in Section \ref{case study 1} (see Eq.\ \rf{tert}), 
where there was no primary constraint $\Pi_\al$, nor a secondary constraint $\Ph_\al$. 
Secondly, 
the lagrange multiplier $v$ appears in the evolution equation for the
$\Ph_\al$ constraint \rf{Phialdot2}.
This latter result also differs from case study 1, 
where $v$ did not even occur because there was no $\Ph_\al$ constraint, 
and in GR, 
$v$ remains an undetermined Lagrange multiplier.

In this case, 
the standard procedure is to solve for $v$ from \rf{Phialdot2} by demanding that 
the equation weakly vanish.
The first three terms vanish weakly by the prior secondary constraints, 
so this amounts to the $v$ term cancelling all remaining terms.
When we solve for $v$ in this manner we obtain
\bea
v &=& -\fr{3 \al \dot{\Om}}{8 (1-\sl )} 
- \fr{\ka \al^2 \Pi}{(1-\sl)\sqrt{\ga}} 
+ \be^i \cD_i \al
\nonumber\\
&&
-\frac { 4 \ka^2 \al^3 (\dot{s} + \dsl ) } {3 \ga (1-\sl )\dot{\Om}^2} 
\left(\Pi_{ij}\Pi^{ij} -\frac{1}{2}\Pi^2 \right)
\nonumber\\
&&
+\left((1+ \tfrac 13 s) \dot{\Om}
- \tfrac 43 \dot{s} (1-\sl ) \right) \fr{\al^3 \cR}{3 \dot{\Om}^2}
\nonumber\\
&&
+\fr{\al \ddot{\Om}}{\dot{\Om}}
+\fr{\al \dsl}{2(1-\sl)}.
\label{v}
\eea
As can be seen from this equation, 
there is a problematic denominator in some of the terms.
One would thus demand that solution only include cases where $\dot {\Om} \neq 0$.

Denoting the solution of \rf{v} with capital $V$, 
the expression would then be inserted back into the Hamiltonian and
the final form would be
\bea
H_F &=& \int d^3 x (\cH_0 + V (\al,\ga_{ij},\be^i,\Pi^{ij},...) \Pi_\al 
+ \xi^i \Pi_i
\nonumber\\
&&
\pt{\int d^3 x }
+ \ze^j \Ph_j ),
\label{HF2}
\eea
where $\cH_0$ \rf{H02} is evaluated with $\prt_i \Om=0$,
and we have indicated that $V$ is now a function of the 
canonical variables and the coefficients.
We have added three additional Lagrange multipliers $\ze^j$
for the secondary constraints $\Ph_j \approx 0$. 
Note that, 
upon doing this, 
we end up with one of the Hamilton's equations specifying 
$\dot{\al} = V(\al,\ga_{ij},\be,\Pi^{ij},...)$, 
again in contrast to GR where $\al$ is pure gauge.
The full Hamilton's equations for this case are lengthy and omitted here, 
but it would be of interest in future work to study these types of cases 
in more detail.

In the result, 
equation \rf{HF2}, 
we have $4$ primary constraints \rf{PC2} and \rf{PA2}, 
$4$ secondary constraints \rf{secondary2a} and \rf{secondary2b}, 
and a total of $6$ Lagrange multipliers $\xi^j$ and $\ze^j$.  
Note that the Lagrange multiplier $v$ was solved for, 
and so does not count as an undetermined
Lagrange multiplier.
Using the counting scheme in equation \rf{dof}, 
for this case we obtain
$10 - (1/2)(8)-(1/2)6 = 3$ degrees of freedom, 
one more than GR.

Another choice is to set $s$ and $\sl$ to be constants.
This choice reduces the Hamiltonian to one where there are 
scalings of GR terms, 
obtainable from \rf{H02} by setting the 
$\Om^\prime, \prt_i \Om$ terms to zero.
Indeed it is this choice that forms the starting point for
the match of explicit breaking models to the SME, 
as we discuss in Section \ref{horavamatch}.
For this latter choice, 
the number of degrees of freedom reduces
to the GR result of $2$.

\subsection{Addition of Matter}
\label{sec:matter}

To apply the results above to physically relevant situations, 
we address the addition of the matter sector to the Hamiltonian
analysis.
We assume here that the matter sector does not couple to any 
coefficients for Lorentz violation
and is minimally coupled to gravity.
Depending on the area of study, 
the description of matter could be as basic as a perfect fluid
or a set of scalar fields, 
or more sophisticated with gauge fields and/or spinors.
For this work we shall leave this specification generic and comment
on how the matter sector feeds into the analysis above.

First note that when performing variations of the matter action with respect to
the spacetime metric $g_\mn$, 
we have 
\beq
(T_M)^\mn = \frac {2} {\sqrt{-g}} \frac {\de S_M}{\de g_\mn}.
\label{SEdef}
\eeq
Upon constructing the Hamiltonian for the matter sector, 
we can use \rf{SEdef} and the 3+1 decomposition to show that 
the following hold in space and time components:
\bea
\frac {\de H_M} {\de \al} &=& \al^2 \sqrt{\ga} (T_M)^{00},
\nonumber\\
\frac {\de H_M} {\de \be^i} &=& 
- \al \sqrt{\ga}[ (T_M)^{00} \be_i + (T_M)^{0j}\ga_{ij} ],
\nonumber\\
\frac {\de H_M}{\de \ga_{ij}} &=& 
- \tfrac 12 \al \sqrt{\ga} 
[ (T_M)^{ij} + \be^i \be^j (T_M)^{00} + 2 (T_M)^{0(i} \be^{j)} ].\nonumber\\
\label{HMvar}
\eea

In the Dirac-Hamiltonian analysis, 
one checks the consistency or evolution 
of the secondary constraints.
If you add the matter sector, 
minimally coupled to gravity, 
certain combinations of the terms
in \rf{HMvar} are involved in these calculations.
For example, 
in the secondary constraints in \rf{secondary2b}, 
an extra term for the matter sector $-\de H_M/\de \be^i$ is added, 
and its evolution is governed by the expression
\bea
\left\{ \frac {\de H_M}{\de \be^i}, H_A \right\} &=& 
\frac {\de H_M}{\de \al} \cD_i \al 
+\cD_j \left( \be^j \frac {\de H_M} {\de \be^j} \right)
\nonumber\\
&&
+ \frac {\de H_M}{\de \be^j} (\cD_i \be^j )
-2 \ga_{ki} \cD_j \frac {\de H_M}{\de \ga_{jk}}
\label{matterEvol}
\eea
where $H_A$ is the augmented Hamiltonian
including the matter sector.
Similar results hold for $\de H/\de \al$.
Finally we note that, 
while we don't address it here, 
the addition of even minimally coupled tensor fields
can significantly affect the constraint structure of the model, 
as shown in Ref.\ \cite{IN77}.

\section{Cosmological Solutions}
\label{sec:cos}

In this section we apply the Hamilton's equations
for the case study 1 subset of the SME discussed in \ref{case study 1}
to search for solutions in a Friedmann-Lemaitre-Robertson-Walker (FLRW) spacetime
\cite{mtw}.
We use the general FLRW metric in spherical coordinates
\beq
    ds^2 = - dt^2 +a^2(t)\Big[ \frac{dr^2}{1-kr^2}+r^2\big(d\th^2+\sin^2{\th}d\ph^2 \big)
    \Big],
    \label{flrw}
\eeq
where $a(t)$ is the scale factor and $k=\{+1,0,-1\}$ represents a closed, flat, 
and open universe, respectively.
For this metric, 
the lapse and shift can be seen by comparison with \rf{ADMmetric} 
to be $\alpha=1, \beta^i=0$
and the acceleration vanishes $a_i =0$.

We proceed with evaluating Hamilton's equations for this case.
Using the result $\dot{\ga}_{ij} = 2 \dot{a} \ga_{ij} /a$,
equations \rf{piG} and \rf{PA} we can find the canonical momenta to be
\bea
    \Pi^{ij} &=&-\frac{\sqrt{\ga}}{\ka}(1-s_{00})\frac{\dot{a}}{a}\ga^{ij}
    +\frac{\sqrt{\ga}}{4\ka} \dot{s}_{00} \ga^{ij},
    \nonumber\\
    \Pi_\al &=& -\frac{ 3s_{00}}{\ka} \sqrt{\ga}\frac{\dot{a}}{a},
    \label{set1}
\eea
where $\sqrt{\ga}=a^3r^2\sin{\th}/\sqrt{1-kr^2}$. 
These results allow us to establish that since $\Pi^{ij}$ is proportional to $\ga^{ij}$, 
quantities like $\Pi^{ij} \Pi_{ij} - \frac 13 \Pi^2$ will vanish.
Indeed, 
evaluation of the other set of Hamilton's equations 
(\ref{eq:eom_alpha}-\ref{eq:eom_gamma}) for this case yields
\bea
    \nonumber\dot{\Pi}^{ij} &=& 
    \frac{\ka (1- s_{00})}{6 \sqrt{\ga} s_{00}^2} \ga^{ij} \Pi_\al^2
    - \frac{\ka }{3 \sqrt{\ga} s_{00}} \Pi_\al (\Pi \ga^{ij} - 2\Pi^{ij} ) 
    \nonumber\\
    &&
    -\frac{\sqrt{\ga}}{2\ka} \cG^{ij} + \frac {\sqrt{\ga}}{2} (T_M)^{ij},     
    \nonumber\\
    \dot{\Pi}_\al &=& 
    \frac{\sqrt{\ga}}{2\ka} \mathcal{R}    
    + \frac{8 \ka }{3\sqrt{\ga}s_{00}} \Pi_\al \Pi  
    - \frac{\ka}{3\sqrt{\ga}s^2_{00} } 
    \left(7 - 5s_{00} \right) \Pi_\al^2 
    \nonumber\\
    &&
     - \frac{1}{2 s_{00}} \Pi_\al \dot{s}_{00} 
     -\sqrt{\ga} (T_M)^{00},
    \label{set2}
\eea
where we have used the matter couplings in \rf{HMvar}
and $\cG^{ij}$ is the three-dimensional Einstein tensor.
With the choice of metric and $\be^j=0$
the constraint equation \rf{eq:eom_beta} is satisfied, 
as can be checked directly.
Also, 
note that the fact that $\al$ is dynamical in our model, 
and even though it is fixed to unity,
it still plays a role through the momenta $\Pi_\al$.

For matter we use the usual perfect fluid model for a homogeneous 
and isotropic universe,
$(T_M)^\mu_{\pt{\mu}\nu} = {\rm diag}(-\rho, p,p,p)$, 
where $\rho$ and $p$ are the energy density and pressure, 
respectively. 
They are related through the equation of state $p= w \rho$, where $w$ is the barotropic index. 

The three-dimensional Ricci scalar and the three-dimensional Einstein tensor
are $\cR = 6 k/a^2$, and $\cG^{ij} = -k \ga^{ij}/a^2$.
Combining with \rf{set1} and \rf{set2}, 
we obtain two equations:
\bea
     \left(\frac{\dot{a}}{a}\right)^2 (1-s_{00}) 
     &=& \frac {\ka \rh}{3} - \frac{k}{a^2} 
     -s_{00} \frac{\ddot{a}}{a} + \frac {\dot{a}}{a} \frac {\dot{s}_{00}}{2},\nonumber\\
     \label{f1}\\
     \left[ \frac {\ddot{a}}{a} + \frac 12 \left(\frac{\dot{a}}{a}\right)^2 \right] (1-s_{00})
     &=& -\frac {\ka p}{2} - \frac{k}{2a^2}  + \frac {\dot{a}}{a} \dot{s}_{00}
     + \tfrac 14 \ddot{s}_{00},\nonumber\\
    \label{f2}
\eea
which have been written to match the standard FLRW equations of 
GR as closely as possible.
Indeed one recovers GR in the limit that $s_{00} \rightarrow 0$.
The modifications include terms with first and second time derivatives of $s_{00}$,
scalings by $1-s_{00}$, 
and an extra $\ddot{a}$ term in the first equation.
In principle, 
one could decouple the equations to obtain
one with only the acceleration $\ddot{a}$ and one with only
the Hubble factor $\dot{a}$, 
in the standard Friedmann equation form.
However, 
$s_{00}$ is an, 
as yet unspecified function associated with explicit breaking 
of the underlying symmetries in the action \rf{lag1}.

In order to understand the role of $s_{00}$ 
in this context more plainly, 
we examine the remaining conservation laws
implied by the underlying action.
These were given in equation \rf{SMEcl} and for this particular subset
of the SME, 
in equation \rf{eq:bianchivacuum}.
Eq.\ \rf{SMEcl} must be satisfied for consistency:
\beq
\nabla_\mu (T_{\rm s})^\mn =- \ka \nabla_\mu (T_M)^\mn.
\label{SMEc1repeat}
\eeq
The $\nu=j$ component has already be satisfied by the assumption in \rf{s00Cond}, 
and thus we can assume correspondingly that 
part of the usual matter conservation law is 
satisfied, $\nabla_\mu (T_M)^\mu_{\pt{\mu}j}=0$.
The Hamiltonian method did not directly involve the $\nu=0$ component, 
so we must ensure that it holds as well in the cosmological solutions here.

After some computation, 
we obtain for the left-hand side of \rf{SMEc1repeat},
\beq
\nabla_\mu (T_s)^\mu_{\pt{\mu} 0} =   \frac {\ddot{a}}{a} 
\left( \tfrac 32 \dot{s}_{00} + 6 s_{00} \frac{\dot{a}}{a} \right)
+3 s_{00} \frac {\dddot{a} }{a}.
\label{FLRWcons0}
\eeq
For the matter part, we obtain
\beq
    \nabla_\mu (T_M)^\mu_{\pt{\mu} 0} = -\dot{\rho} - 3\frac{\dot{a}}{a}\left(\rho+p\right).
    \label{FLRWconsMatter0}
\eeq
Consistency of these results, 
i.e., left-hand side equals right-hand side, 
can be verified with the equations \rf{f1} and \rf{f2} 
by solving for $\rh$ and $p$ and inserting 
the expressions into \rf{FLRWconsMatter0}, 
to recover \rf{FLRWcons0}.

We are now in a position to examine the consequences
of different choices for $s_{00}$.
Among the myriad of possible functional forms
for $s_{00}$, 
we study two cases here.
First, 
we look at the case where $s_{00}$ is determined by
demanding that the matter stress-energy tensor by itself 
is completely conserved and thus equation \rf{FLRWconsMatter0} vanishes.
Second, 
we look at a case when equation \rf{FLRWconsMatter0} does not vanish, 
yet the total conservation law \rf{SMEc1repeat} holds.

\subsection{FLRW Example 1}
\label{FLRWsoln1}

For the first case, 
we enforce the matter energy-momentum conservation law.
Note that if the matter equations of motion are satisfied (``on shell")
this condition would necessarily hold.
This condition implies that $\nabla_\mu (T_s)^\mu_{\pt{\mu} 0} = 0$
and thus the following expression must be solved for $s_{00}$:
\beq
-\frac {\ddot{a}}{a} 
\left( \tfrac 32 \dot{s}_{00} + 6 s_{00} \frac{\dot{a}}{a} \right)
= 3 s_{00} \frac {\dddot{a} }{a}.
\label{s00eqn}
\eeq
It turns out that an analytical solution for $s_{00}$ given by
\beq
s_{00} = \frac {\ze}{a^4 \ddot{a}^2 }, 
\label{s00soln}
\eeq
solves this equation, 
where $\ze$ is an arbitrary constant.
This solution has intriguing and yet pathological features.
Obviously, 
if the acceleration $\ddot{a}=0$, 
as it does in the past for standard cosmological solutions,
it diverges.
Also, 
if we could assume constant behavior for $\ddot{a}$, 
then the result shows that the coefficient $s_{00}$ would naturally
decrease with the expanding universe.

The next step to pursue the case just outlined would be
to insert the solution \rf{s00soln} back into the modified
equations \rf{f1} and \rf{f2} and attempt to solve the 
resulting system of equations for $a(t)$ for different
choices of sources $\rh$ and $p$.
However, 
one finds that the resulting equations have up to $4th$ order
time derivatives in them, 
if they are solved with no approximations made.
Furthermore, 
it is challenging to approach the equations from
a perturbative point of view, 
where the dimensionless $s_{00}$ is ``small" compared to unity, 
as can be seem from plugging in a GR solution into \rf{s00soln}:
again when $\ddot{a}$ approaching zero, 
this grows large and conflicts with perturbation theory.
In this work, 
we do not pursue this solution further, 
and leave it as an open problem to explore.

\subsection{FLRW Example 2}
\label{FLRWsoln2}

We now turn to the case where we do not impose the vanishing of 
$\nabla_\mu (T_s)^\mu_{\pt{\mu} 0}$.
The coefficient $s_{00}$ remains arbitrary and so for this work 
we examine the simplest case of a constant coefficient, 
$\dot{s}_{00}=0$.
As a consequence of this choice, 
the matter conservation law gets modified
and matter exhibits a modified cosmological evolution
in the presence of $s_{00} \neq 0$.
First we write the Friedmann equations 
for this case as
\bea
\left(\frac{\dot{a}}{a}\right)^2 &=& 
\frac {\ka \rh}{3 (1-\tfrac 32 s_{00} )}
- \frac {k}{ a^2 (1-s_{00}) } 
\nonumber\\
&&+ \frac {\ka p s_{00} }{(2-3 s_{00})(1-s_{00})},
\nonumber\\
\frac{\ddot{a}}{a} &=& - \frac {\ka (\rh + 3 p)}{6 (1-\tfrac 32 s_{00} )}.
\label{feqns:set2}
\eea
These results contain various scalings of the usual GR terms but also
a nonstandard appearance of the pressure in the first equation.

Using \rf{feqns:set2} and \rf{FLRWconsMatter0}
we obtain the modified conservation law, 
or continuity equation,
as 
\beq
    \dot{\rho}+3\frac{\dot{a}}{a}f(w,s_{00})\rho = 0,
    \label{eq:cont}
\eeq
where we have introduced the auxiliary equation $f(w,s_{00})$ as
\beq
    f(w,s_{00}) = \frac{2(1 + w -s_{00})} {2 +s_{00}(3w -2)},
    \label{f}
\eeq
which reduces to the proper GR limit, 
where $f \to 1+w$ as $s_{00} \to 0$. 
We integrate the modified continuity equation to find that
\beq
    \rh=\rh_0\left(\frac{a}{a_0}\right)^{-3 f(w,s_{00})},
    \label{rhmod}
\eeq
where $a_0$ is the present value of
the scale factor.
For matter as a dust $w=0$ and $f=1$ so 
there is no modification to the cosmological evolution 
$\rh \sim a^{-3}$.
However, 
for radiation $w=1/3$ and the cosmological constant $w=-1$,
the evolution equation is modified, 
as occurs in other modifications to GR \cite{gasp85,DynDE}.

This leads to interesting type-dependent evolution of the different 
cosmological fluids. 
We can put together the Friedmann equations, 
paralleling the usual methods, 
by using dimensionless density parameters $\Omega_X$, where $X=\{m,r,\Lambda,k\}$ represents the energy density of matter, radiation, cosmological constant, and curvature, respectively.
We divide the first of equations \rf{feqns:set2} by the square of 
the present value of the Hubble constant $H_0^2 = \dot{a}_0^2/a_0^2$ and use 
the evolution equation \rf{rhmod}.
The result can be written
\beq
   \frac {H^2}{H_0^2} = \Om_{m0} a^{-3} + \Om_{r0} a^{-4\et_r} 
   +\Om_{\La 0} a^{\et_\La} +\Om_{k0} a^{-2},
    \label{F1norm}
\eeq
where $H=\dot{a}/a$, 
$\et_r = (1-\tfrac 34 s_{00})/(1-\tfrac 12 s_{00} )$, 
and $\et_\La=3 s_{00}/(1-\tfrac 52 s_{00} )$.
Note that matter ($m$) and curvature ($k$) behave normally while
radiation ($r$) and the cosmological constant ($\La$) differ from GR, and that the density parameters in \rf{F1norm}
are evaluated at the present epoch $t_0$, as indicated by the subscript 0.
The density parameters here can be found 
for each universe constituent from
\beq
\Om_X = \fr {\ka \rh }{3 H^2 (1-\tfrac 32 s_{00} )} 
\frac {2 + (3w-2) s_{00}}{2 (1-s_{00})}, 
\label{omegas}
\eeq
and for curvature $\Om_k = -k/[H^2(1-s_{00})]$.  
Note that scalings by $s_{00}$ have been absorbed 
into the definitions of the $\Om_X$ values.

Next we examine the acceleration equation for this case.
Using the same density parameters, 
the second of equations \rf{feqns:set2} can be written
\bea
\frac {\ddot{a}}{a H_0^2} &=& -\tfrac 12 \Om_{m0} a^{-3} 
-\Om_{r0} \frac {2 (1-s_{00})}{2-s_{00}} a^{-4\et_r} 
\nonumber\\
&&
+\Om_{\La 0} \frac {2 (1-s_{00})}{2-5s_{00}} a^{\et_\La}.
\label{F2norm}
\eea
Here we can see that the scalings appearing cannot be completely 
removed by re-defining constants.

\begin{figure}[h]
\begin{center}
\vspace{2mm}
\hspace{-7mm}
    \includegraphics[width=0.5\textwidth]{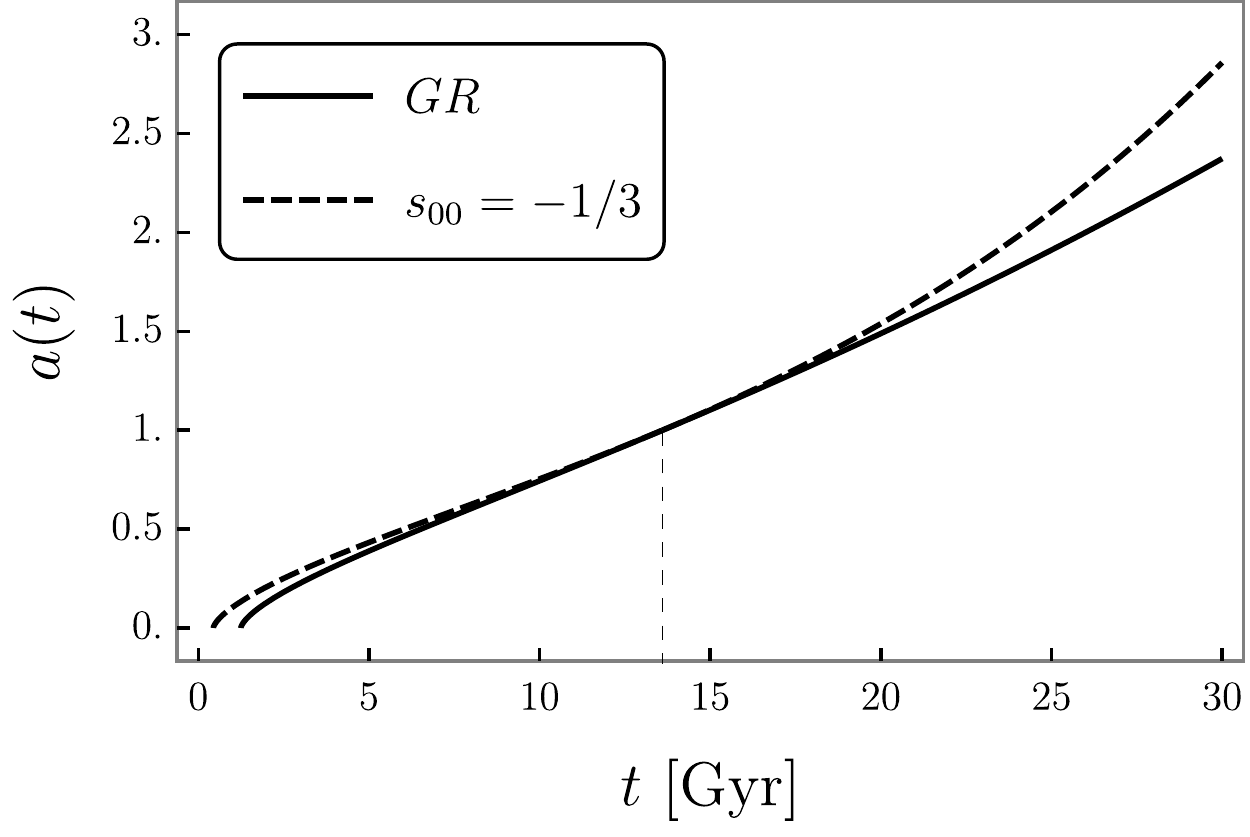}
    \caption{Evolution of the scale factor for
    the constant $s_{00}$ case of the flat FLRW solutions compared to GR, 
    assuming $\Om_{r0}=0$, $\Om_{m0}=0.31$, and $\Om_{\La 0}=0.69$. The dashed vertical line represents the present day.}
    \label{fig2}
    \end{center}
\end{figure}

Equation \rf{F2norm} gives us the deceleration parameter, 
$q\equiv -(\ddot{a}/a)H^{-2}$, 
which we can attempt to use to find a crude constraint on $s_{00}$. 
Since the value of $q$ at the present epoch ($t=t_0$) needs to be negative 
in order to match the observed accelerated expansion we can conservatively
write the inequality
\beq
    -\tfrac 12 \Om_{m0} 
    -\Om_{r0} \frac {2 (1-s_{00})}{2-s_{00}}
    +\Om_{\La 0} \frac {2 (1-s_{00})}{2-5s_{00}} > 0.
    \label{bound}
\eeq
Other than showing that $s_{00}$ is less than order unity, 
this result is not particularly useful for placing constraints, 
since it is challenging to disentangle the density parameters
from the $s_{00}$ coefficient.
Thus a complete analysis using cosmological data \cite{cosmo} 
should be attempted in the future.
To display what the effects of the modified evolution would look like, 
we solve the first Friedmann equation in \rf{F1norm} and plot
in Figure \ref{fig2}.

\section{Connection to models and frameworks}
\label{sec:match}

The SME is a test framework and as such, 
any action-based model that describes coordinate-independent
Lorentz violation, 
should in principle be contained in some subset of terms.
In practice, 
this can be challenging when certain assumptions
are made in the SME to afford tractable phenomenological analyses \cite{bk06,ms09}, 
while these assumptions can differ from those made in specific models.
We show here first how the results in this paper
match to prior work in linearized gravity, 
and then we find a match to models formulated in the 3+1 formalism.

\subsection{Quadratic SME gravity sector}
\label{sec:quadSME}

In references \cite{bk06, wfanalysis, km16, km18}, 
results in the linearized gravity limit have been developed.
In particular, 
a classification of all possible Lorentz-breaking 
Lagrangian terms at quadratic order in the metric fluctuations 
$h_\mn$ around a flat background has been performed.
Such terms take form $\cL \sim h_\mn \hat{{\cal K}}^{\mu\nu\rh\si} h_{\rh\si}$
and much phenomenological analysis already exists, 
including results at leading order in the coefficients in propagation studies.
In linearized gravity, 
diffeomorphism invariance can be described using the gauge transformation
of the metric fluctuations $h_\mn \rightarrow h_\mn - \prt_\mu \xi_\nu -\prt_\nu \xi_\mu$.
The analysis of the quadratic action terms includes both gauge symmetry breaking, 
and gauge symmetric terms, 
though scant phenomenological attention has been put on the former.

We seek here to match the explicit breaking limit of the SME that 
we have used in this work to a subset of these terms
in the weak field limit.  
This will illuminate how results in this work may match
to those previously obtained. 
Curiously, 
the SME Lagrangian with $u$, $s_\mn$, 
and $t_\abgd$ terms can be shown to have both gauge-breaking and gauge-symmetric terms
in the quadratic action limit when taken in the explicit breaking limit. 
Furthermore, 
one can trace the occurrence of dynamical pieces of the metric fluctuations 
$h_\mn$ that are non-dynamical in GR and in gauge-symmetric models. 
To see this we first examine the contributions of 
the $s_\mn$-type term only:
\bea
\cL_s &=& \frac{\sqrt{-g}}{2 \ka} s_\mn R^\mn 
\nonumber\\
 &=& \frac{\sqrt{-g}}{2 \ka} 
 \left( s_\mn G^\mn + \fr 12 s^\la_{\pt{\la}\la} R \right), 
\label{Ls}
\eea
where no linear approximations have yet been made.

Next we assume a weak-field expansion around a flat background for both the metric
and $s_\mn$:
\bea
g_\ab &=& \et_\ab + h_\ab,
\nonumber\\
s_\ab &=& \sb_\ab + \st_\ab.
\label{exp}
\eea
We keep fluctuations for $s_\ab$ for generality 
at this point and we will assume that the partial derivatives of $\sb_\mn$ vanish.
The Lagrange density \rf{Ls} is then expanded in the quadratic action limit 
(keeping terms of order $h^2$, $h \st$, $\st^2$ and discarding total derivatives).
It can be then be written as
\bea
\cL_s &\approx&  
\frac {1}{2\ka} \Big[ (1+ \tfrac 12 h) \sb_\ab G^\ab + \st_\ab (G_L)^\ab
\nonumber\\
&&
\pt{\ka} 
+\tfrac 12 (1+\tfrac 12 h) \et^\mn \sb_\mn R 
+ \tfrac 12 (\et^\mn \st_\mn -h^\mn \sb_\mn ) R_L \Big],\nonumber\\
\label{LsQ}
\eea
where curvature terms with the subscript $L$ are linearized, 
and those without are taken to quadratic order.
It turns out that the first term on the first line of \rf{LsQ} by itself reproduces
the gauge invariant contribution to the SME quadratic action expansion for the $\sb_\mn$ term,
\beq
\frac {1}{2\ka} (1+ \tfrac 12 h) \sb_\ab G^\ab = \frac {1}{4\ka} \sb_\ab h_\gd \cG^{\al\ga\be\de},
\label{sme}
\eeq
where $\cG^{\al\ga\be\de}$ is the linearized double dual curvature tensor \cite{bCPT10}.
Thus, 
if we take the explicit-breaking limit by discarding
the fluctuations $\st_\mn$ entirely, 
we end up with the sum of the gauge invariant quadratic action terms 
and gauge-violating terms.

To summarize so far: in the quadratic action limit
\beq
\cL_{s, \rm explicit} = \frac {1}{4\ka} \left(\sb_\ab h_\gd \cG^{\al\ga\be\de} 
- h^\mn \sb_\mn R_L \right), 
\label{LsQ2}
\eeq
where the second term is explicitly gauge-violating 
and can be matched to the general expansion of Ref.~\cite{km18}, 
and we have discarded the trace $\et^\mn s_\mn$ term 
that merely scales GR.
Among the gauge-violating terms in \cite{km18}, 
at mass dimension $4$, there are two types of terms 
which are relevant for the second term in \rf{LsQ2}. 
They are contained in the general expansion in Table 1 of \cite{km18}:
\bea
h_\mn \hat{{\cal K}}^{\mu\nu\rh\si} h_{\rh\si} \supset 
h_\mn (s^{(4,1)\mu\rh\nu\si\al\be} + k^{(4,1)\mu\rh\nu\si\al\be}) \prt_\al \prt_\be h_{\rh\si},\nonumber\\
\label{gvterms}
\eea
where the $\mu\rh\nu\si$ indices are totally symmetric in the $s^{4,1}$ coefficients
and of Riemann tensor symmetry $[\mu\rh][\nu\si]$ for the $k^{4,1}$ coefficients. 
The match to these terms for the present case can be obtained using the form
\bea
h^\mn \sb_\mn R_L = \frac {1}{2} 
h_\mn ( \sb^\mn \hat{K}^{\rh\si} + \sb^{\rh\si} \hat{K}^\mn )h_{\rh\si},
\label{gvtermsmatch}
\eea
where $\hat{K}^\mn = \prt^\mu \prt^\nu - \et^\mn \prt^\la \prt_\la$.
To complete the match one has to take the 
appropriate symmetric and antisymmetric combinations
of the quantity in parentheses in \rf{gvtermsmatch}.

Finally, 
we note that the fact that the terms studied in this paper 
correspond to the gauge-violating limit of the SME quadratic expansion
explains, in part, 
why additional degrees of freedom beyond GR are found, 
as in Section \ref{case study 1}.
For instance, 
because of the symmetries of the operator $\hat{{\cal K}}^{\mu\nu\rh\si}$
for gauge-symmetric terms, 
it can be shown that no time derivatives of $h_{00}$ appear 
when the Lagrange density is written in the first-order derivative form 
$\sim \prt h {\cal K} \prt h$.
Any such terms would correspond to time derivatives 
of the lapse function $\al$ via $\al\approx 1+ h_{00}/2$ in the weak-field limit. 
For gauge-violating terms, 
as in equation \rf{gvtermsmatch}, 
such terms can appear because the symmetries 
of the operators $\hat{{\cal K}}^{\mu\nu\rh\si}$ allow for them, 
as they are less restrictive.
In the case of $\sb_{00}$ being the only nonzero coefficient we have 
\bea
\cL_s &\supset& \frac{1}{4\ka} \sb_{00} 
[ \prt_0 h_{00} (\prt_0 h_{jj} - \tfrac 12 \prt_j h_{j0} ) + ...]
\label{h00dot}
\eea

Despite this interesting feature there are likely severe constraints 
on any such models via the traced Bianchi identities, 
even in the linearized gravity limit.
For example, 
for the case \rf{LsQ2}, 
the field equations from the first term are gauge invariant 
and automatically satisfy the traced Bianchi identities.
The second term however, 
would yield a constraint in the presence of matter given by
\beq
\frac 12 \prt_\mu (\sb^\mn R_L)=\ka \prt_\mu (T_M)^\mn.
\label{constraint}
\eeq
Thus one either has a Ricci flat restriction which is challenging 
to reconcile in the presence of matter, 
or one has a modified conservation law for matter, 
or one must reject such cases (``no-go").
We showed in Section \ref{sec:cos} that modified behaviour 
of matter may be an acceptable solution in some cases, 
like cosmology.

\subsection{Match to 3+1 models}
\label{horavamatch}

Matches of specific models of Lorentz violation
to the SME has been accomplished in the gravitational sector 
for a variety of models including those with dynamical vectors and tensors, 
noncommutative geometry, 
and massive gravity models \cite{bbw19}.
Among the proposals for renormalizable quantum gravity is the approach known 
as Ho\v{r}ava-Lifshitz (HL) gravity \cite{Horava}.
Since this model is based on a 3+1 formalism 
we should be able to match it the SME in the present work.
We shall focus on a simpler version of this model 
where the action is written in the 3+1 form:
\bea
\cL_H = \al \sqrt{\ga} (K_{ij} K^{ij} - \la K^2 + \xi \cR + \et a^i a_i + \hdots) 
\label{LH}
\eea
where the ellipses include possible higher order spatial derivative terms 
and the matter sector \cite{blas09, frvs16}.
(For simplicity in the remainder of this section we set the coupling $2\ka=1$.)

Note that the insertion of a parameter in front of the terms that occur 
in GR is akin to early kinematic approaches of tests of special relativity 
and dressed-metric based approaches for tests of GR \cite{cmw}.
That approach seems somewhat ad-hoc from the SME point of view, 
since the SME is based on observer covariant terms added to the action
with coefficients with indices. 
Nonetheless, 
we can possibly accommodate these terms with certain components 
of the SME coefficients in a particular coordinate system, 
as has been done for other models \cite{kmPhot}. 
Eq.~\rf{LH} above 
takes a rotationally isotropic form.
If we proceed with the $s_\mn R^\mn$ coupling in the isotropic limit 
presented in \ref{case study 2}, 
assuming the coefficients are constant in time and space, 
we obtain 
\beq
    \cL_2 = \al \sqrt{\ga}
    \Big[
    \cR \left(1+ \tfrac 13 s \right) + \left(K^{ij}K_{ij} -K^2 \right)(1-\sl)
    \Big]. \\
\label{Case2H}
\eeq
Note that in the isotropic limit the combination 
$K^{ij} K_{ij} -K^2$ cannot be broken apart with an $s_\mn$-type term alone;
however, 
in the conception of the SME as a limit of spontaneous symmetry breaking 
we have the freedom to add dynamical terms to the action.
For example, 
for the $s_\mn$ coefficients we can add general dynamical terms \cite{b19}, 
which are included in the Appendix \ref{appDyn}, 
to match \rf{LH}.

We take first the case where $\sl=0$ in \rf{Case2H} 
and add the terms labeled $5$ and $12$ in the Appendix
with a distinct set of coefficients that we denote
with a capital $S_\mn$. 
This yields
\bea
\cL_{\rm{SME, Match}} &=& \al \sqrt{\ga} 
\big[ \cR \left(1+ \tfrac 13 s \right) + K^{ij}K_{ij} -K^2
\nonumber\\
&&
+a_5 \tfrac 12 (\nabla_\mu S^{\mu\la}) (\nabla_\nu S^\nu_{\pt{\be}\la} )
\nonumber\\
&&
+a_{12} (S^\mn \nabla_\mu S_{\nu\la}) 
(S^{\ka\rh} \nabla_\ka S_\rh^{\pt{\rh}\la})
\big].
\label{match1}
\eea
All terms in this lagrange density are now treated as non-dynamical.
We next assume for the last two terms 
that the only nonzero coefficient in the local frame
is $S_{\bar{0}\bar{0}}=1$ - note the precise value of the coefficient needed.
Using the vierbein \rf{ADMvierbein} one can show that this 
is equivalent to $S_\mn=n_\mu n_\nu$.
This kind of choice has been used to match HL gravity 
to vector models \cite{jHorava}. 
With these assumptions we arrive at
\bea
\cL_{\rm{SME, Match}} &=& \al \sqrt{\ga} 
\big[ \cR \left(1+ \tfrac 13 s \right) + K^{ij}K_{ij} 
\nonumber\\
&&
-K^2 (1+\tfrac 12 a_5 )
+ \left(a_{12} + \tfrac 12 a_5 \right) a^i a_i \big].
\nonumber\\
\label{match2}
\eea
It is now clear that if we make the following choice, 
$\la = 1+ a_5/2$, $\xi=1+s/3$, 
and $\et=a_{12}+a_5 /2$, 
then HL gravity in the form \rf{LH} can be matched to this limit of the SME.
Note that the extra terms added to the SME are of second order and fourth order in $S_\mn$, 
and two distinct sets of coefficients were used in this match.
Finally,
while we do not discuss it here, 
matter couplings proposed in the literature have also been matched 
to the matter sector of the SME in Ref.\ \cite{bbw19}.

\section{Discussion \& Conclusion}\label{sec:CON}

In this work, 
we have taken initial steps towards exploring the SME effective field theory framework
description of local Lorentz and diffeomorphism breaking in the
areas of the 3+1 formalism, 
Dirac-Hamilton analysis of the dynamics, 
and cosmology.
We have examined consequences of adopting 
the explicit symmetry breaking paradigm, 
which is complementary to existing work assuming spontaneous symmetry breaking.
Furthermore, 
we have established results without using 
the weak-field gravity approximation.

The key results of this work include a 3+1 decomposition 
of the SME gravity sector actions 
in Section \ref{GR and the SME action}, 
including a general analysis of the time derivative terms 
that occur, 
relevant for Hamiltonian analysis.
We studied two example subsets of the SME
using the Dirac-Hamiltonian analysis in 
Section \ref{sec:ham}.
The results of one of these cases, 
the Hamilton's equations in (\ref{eq:eom_alpha}-\ref{eom:qs}), 
were studied for FLRW cosmological solutions in Section \ref{sec:cos}, 
where some novel cosmological evolution was found.
Further analysis for other strong-field gravity solutions 
can be the subject of future work, 
for instance black hole spacetimes or other exotic solutions \cite{worm}.
We also established a link between the explicit breaking terms
in this work and existing SME studies in linearized gravity
and we further elucidated the match to Ho\v{r}ava-Lifshitz gravity in
Section \ref{sec:match}.

A set of Hamilton's equations like those found in Section \ref{sec:ham}
for a subset of the SME can be used to study the initial value formulation, 
and develop numerical techniques to simulate Lorentz-breaking effects
on strong-field gravitational systems \cite{comp}.
Results in this paper can also be applied to a 3+1 and Dirac-Hamiltonian 
analysis of spontaneous-symmetry breaking scenarios, 
for example by using the second order $s_\mn$ terms in \rf{dynS}.

One of the notable results of this work 
is the identification of subsets of the SME, 
whereupon in the explicit breaking limit, 
extra degrees of freedom, 
normally gauge in GR, 
occur in the Hamiltonian analysis.
In light of this, 
it would be of interest to investigate 
approaches to quantum gravity \cite{deWitt} 
and the role of the 
``problem of time" in the SME framework \cite{pot}.
Also, 
we explored the severe constraints that exist on such a model, 
due constraints imposed by the Bianchi identities 
coming from the underlying geometrical framework \cite{kl20}.
The analysis in this paper represents a first step towards studying this phenomena, 
and it remains an open problem to fully understand
the nature of these extra degrees of freedom.

As a preview of future work, 
we note that the cosmological solutions in Section \ref{sec:cos}
can be obtained from an effective classical Hamiltonian 
for homogeneous spacetimes with the variables $a(t)$, $\al(t)$, 
their conjugate momenta $p_a=\prt L/\prt \dot{a}$ and $p_\al=\prt L/ \prt \dot {\al}$,
and matter variables, 
where $L$ is the classical lagrangian.
This ``mini-superspace" Hamiltonian 
takes the form, 
for vanishing spatial curvature and up to scalings,
\beq
H  = \frac {\ka \al^5 (\al^2 - s_{00} ) p^2_\al}{3 a^3 s^2_{00}} 
- \frac {\ka \al^4 p_\al p_a }{3 a^2 s_{00} } +H_M,
\eeq
with matter Hamiltonian $H_M$.
This would modify the widely-studied Wheeler-deWitt equation \cite{qc}, 
for which $\al$ is nondynamical and a $p_a^2$ term is present instead.
Indeed,
since the usual Hamiltonian constraint is absent in this model, 
the wave function $\Ps=\Ps (a, \al,..., t)$ would depend 
on time $t$ and evolve according to the Schr\"odinger equation 
$i \prt_t \Ps = H \Ps$. 
We expect this could offer a new area of exploration 
in quantum cosmology, 
and will be studied in the future.

\begin{acknowledgements}

We thank R.\ Bluhm, Y.\ Bonder, M.\ Seifert, V.\ Svensson, and A.\ Miroszewski for valuable discussions.  
The contributions of Q.G.B. and K.O.A. were supported in part by the 
National Science Foundation under grant no.\ 1806871. 
N.A.N. was supported by the National Centre for Nuclear Research. 

\end{acknowledgements}

\section{Appendix}

\subsection{3+1 formalism}
\label{sec:app3+1}

In the $3+1$ formalism we can express projections of the curvature tensors 
in terms of the timelike normal to the spatial hypersurfaces $n^\mu$, 
the inverse spatial metric $\ga^\mn$, 
the extrinsic curvature $K_\mn$, spatial covariant derivative $\cD_\mu$, 
the Lie derivative along the normal vector $\cL_{\bf n}$, 
the acceleration $a_\mu$, and the 3 dimensional curvature tensor $\cR_\abgd$.
This decomposition is standard in the literature \cite{adm,mtw}, 
but for completeness we record here some useful results 
that can be derived from existing published ones.
First, the basic relations for the 3+1 projections 
of the 4 dimensional curvature tensor 
are given by
\bea
\ga^\al_{\pt{\al}\mu} \ga^\be_{\pt{\be}\nu} \ga^\ga_{\pt{\de}\ka} \ga^\de_{\pt{\de}\la} R_\abgd
&=& \cR_{\mu\nu\ka\la} + K_{\mu\ka} K_{\nu\la} -K_{\mu\la} K_{\nu\ka}, 
\nonumber\\
\ga^\al_{\pt{\al}\mu} \ga^\be_{\pt{\be}\nu} \ga^\ga_{\pt{\de}\ka} n^\de R_\abgd
&=& \cD_\nu K_{\mu\ka} - \cD_\mu K_{\nu\ka},
\nonumber\\
\ga^\be_{\pt{\be}\mu} \ga^\de_{\pt{\de}\nu} n^\al n^\ga R_\abgd
&=&  \cL_{\bf n} K_\mn + \frac {1}{\al} \cD_\mu \cD_\nu \al 
+ K^\be_{\pt{\be}\mu} K_{\nu \be}.
\nonumber\\
\label{curvproj}
\eea
From these, by taking contractions, we have the following decomposition of the four-dimensional curvature Ricci tensor:
\bea
R^\mn &=& \cR^\mn + n^\mu K^{\nu\al} a_\al +n^\nu K^{\mu\al} a_\al
+ K K^\mn
- \cL_{\bf n} K^\mn 
\nonumber\\
&& + 2 K^{\al\mu} K^\nu_{\pt{\nu}\al}
 -a^\mu a^\nu - \cD^\mu a^\nu 
-n^\mu \cD^\nu K - n^\nu \cD^\mu K
\nonumber\\
&& +n^\mu \cD_\al K^{\al\nu}  + n^\nu \cD_\al K^{\al\mu}
\nonumber\\
&&+ n^\mu n^\nu 
\left( \cL_{\bf n} K - K^\ab K_\ab + a^2 + \cD_\al a^\al \right)
\label{4Ricci}
\eea

It is also useful to have a form for the curvature tensors which includes total spacetime covariant derivatives 
rather than Lie derivatives and spatial covariant derivatives.
Using the definitions and properties of spatial covariant derivatives and Lie derivatives,
Eqs.~\rf{curvproj} can be manipulated to the following forms:
\bea
R &=& \cR + K^\ab K_\ab - K^2 - 2 \nabla_\al (n^\al K + a^\al ),
\nonumber\\
R^\ab  &=& \cR^\ab -2 K^\ab K + 2 K^\al_{\pt{\al}\de} K^{\de\be} - n^\al a^\be K
\nonumber\\
&&
+n^\al K^\be_{\pt{\be}\de} a^\de - n^\al n^\be (K^2- K^\ab K_\ab )
\nonumber\\
&&
+\nabla_\de [ n^\al n^\be (n^\de K + a^\de ) - n^\de K^\ab - \ga^{\de\be} a^\al 
\nonumber\\
&&
-(n^\al \ga^{\be\de} +n^\be \ga^{\al\de}) K + n^\al K^{\be\de} +n^\be K^{\al\de} ],
\nonumber\\
R^\abgd &=& \cR^\abgd -3 (K^{\al\ga} K^{\be\de}-K^{\be\ga} K^{\al\de}) 
\nonumber\\
&&
+ (K^{\al\ep} K^\ga_{\pt{\ga}\ep} n^\be n^\de + {\rm sym})
- (K^{\al\ga} n^\be n^\de K + {\rm sym})
\nonumber\\
&&
- (K^{\al\ga} n^{(\be} a^{\de)} + {\rm sym})
\nonumber\\
&&
+\nabla_\ep \big[ n^\ep (K^{\al\ga} n^\be n^\de + {\rm sym})
\nonumber\\
&&
\pt{+\nabla} 
+ (\ga^{\ep (\al} a^{\ga)} n^\be n^\de + {\rm sym} )
\nonumber\\
&&
\pt{+\nabla} 
-2 ( K^{\al\ga} n^{(\be} \ga^{\de)\ep} + {\rm sym} ) \big]
\label{curvproj2}
\eea
where in the last equation, ``sym" refers to the 
Riemann symmetric combination of terms involving the indices $\al,\be,\ga,\de$.  
For instance, for two symmetric tensors $A^{\al\ga} B^{\be\de}+{\rm sym}=A^{\al\ga} B^{\be\de}-A^{\be\ga} B^{\al\de}-A^{\al\de} B^{\be\ga}+A^{\be\de} B^{\al\ga}$.

Results using the explicit form for the metric \rf{ADMmetric} 
are used throughout this paper, 
and some key expressions are collected here.
The three-dimensional connection coefficients 
are given explicitly in terms of the metric $\ga_{ij}$:
\beq
\3g{i}{j}{k} = \frac 12 \ga^{il} 
(\prt_j \ga_{kl} + \prt_k \ga_{jl} - \prt_l \ga_{jk} ),
\label{3conn}
\eeq
where $\ga^{il}$ is the inverse of the 3 metric and satisfies $\ga^{il} \ga_{lk}=\de^i_{\pt{i}k}$.
The components of the spatial covariant derivative acting on an arbitrary covariant vector $v_\mu$ are given by
\bea
\cD_0 v_0 &=& \be^i \be^j (\prt_i v_j - \3g{k}{i}{j} v_k + n^\mu v_\mu K_{ij}),
\nonumber\\
\cD_0 v_i &=& \be^j (\prt_j v_i - \3g{k}{i}{j} v_k + n^\mu v_\mu K_{ij} ),
\nonumber\\
\cD_i v_0 &=& \be^j (\prt_i v_j - \3g{k}{i}{j} v_k + n^\mu v_\mu K_{ij} ),
\nonumber\\
\cD_i v_j &=& \prt_i v_j - \3g{k}{i}{j} v_k + n^\mu v_\mu K_{ij},
\label{Dv}
\eea
where $n^\mu v_\mu = (1/\al)(v_0 - \be^i v_i)$.

\subsection{Poisson bracket analysis}
\label{appPB}

In this subsection we collect some key results on Poisson brackets 
in field theory for the Dirac-Hamiltonian analysis that we use in the paper.
Some results can be found in various places in the literature 
\cite{boj,ms1} but some subtleties arise in the calculations
and it is useful to record them explicitly here.
Firstly, for fields $q_n (t, \vec r)$, 
momenta $p^n (t, \vec r)$, 
and functions of the fields and momenta $f(q,p)$ and $g(q,p)$, 
the Poisson bracket definition is formally
\beq
\{ f , g \} = \int d^3z 
\left( 
\frac {\de f}{\de q_n (t,\vec z)} \frac {\de g}{\de p^n (t,\vec z)}
- \frac { \de f}{\de p^n (t,\vec z)} \frac {\de g}{\de q_n (t,\vec z)} \right),
\label{PBdef}
\eeq
where $f$ and $g$ may depend on different spatial points 
via their dependence on the fields and momenta.
Note also the equal times for all the fields.
As an example, 
if we examine a single scalar field 
and let $q_1 = \ph (t, \vec r)$ and the conjugate momenta $p_1=\Pi = \Pi (t, \vec r^\prime)$, 
then we obtain:
\beq
\{ \ph (t, \vec r), \Pi (t, \vec r^\prime ) \}
= \de^3 (\vec r - \vec r^\prime ).
\label{PBex}
\eeq

In classical mechanics, 
the functions $f$ and $g$ are algebraic functions of the coordinates and momenta.
In field theory however, one often encounters spatial derivatives in 
the calculations of Hamilton evolution via Poisson brackets.
Generically, for a partial spatial derivative $\prt_i$ of a function $f$ of the canonical variables, 
its Poisson bracket with another function $g$ can be shown to obey
\beq
\{ \prt_i f, g \} = \prt_i \{ f,g \},
\label{PBderiv}
\eeq
where the derivative acts on the space dependence $x^j$ of the result of the bracket of $f$ and $g$.
This result can be extended to covariant spatial derivatives.
For example, for the quantity which occurs in GR and the SME for the momentum constraint
$\cD_i \Pi^i_{\pt{i}k}$,  and its Poisson bracket with the Hamiltonian $H$,
using \rf{PBdef} and \rf{PBderiv} we find
\bea
\{ \ga_{kl} \cD_i \Pi^{il}, H \} &=&  \{ \ga_{kl}, H \} \cD_i \Pi^{il} +
\ga_{kl} \cD_i \{ \Pi^{il}, H \} 
\nonumber\\
&& + \Pi^{ij} \cD_i \{ \ga_{jk}, H \}
-\frac 12 \Pi^{jl} \cD_k \{ \ga_{jl}, H \}. \nonumber\\
\label{PBcovD}
\eea
It is important to note that we used the fact that $\Pi^{ij}$ 
is a 3 dimensional tensor density of weight $-1$
and that the spatial covariant derivative has
a dependence on the spatial metric $\ga_{ij}$, 
resulting in the last two terms.

\subsection{Higher-order terms}
\label{appDyn}

The following terms generalize gravitational couplings to
curvature for the SME for the $s_\mn$ term with scalar dimensionless coupling parameters $a_n$:
\bea
\cL_{\rm{s}} &=& \fr {\sqrt{-g}}{2\ka } \Big[ 
a_1 s^\la_{\pt{\la}\la} R
+a_2 s_\mn R^\mn
\nonumber\\
&&
+a_3\tfrac 12 (\nabla_\mu s_{\nu\la}) (\nabla^\mu s^{\nu\la}) 
+a_4 \tfrac 12 (\nabla_\mu s^{\mu\la}) (\nabla_\la s^\be_{\pt{\be}\be}) 
\nonumber\\
&&
+a_5 \tfrac 12 (\nabla_\mu s^{\mu\la}) (\nabla_\nu s^\nu_{\pt{\be}\la} )
+a_6 \tfrac 12 ( \nabla_\mu s^\nu_{\pt{\be}\nu} ) (\nabla^\mu s^\la_{\pt{\be}\la} )
\nonumber\\
&&
+a_7 s_\mn s_{\ka\la} R^{\mu\ka\nu\la}
+a_8 s_\mn s^{\mu}_{\pt{\mu}\la} R^{\nu\la}
\nonumber\\
&&
+a_9 s^\la_{\pt{\la}\la} s_\mn R^\mn
+a_{10} s^\mn s_\mn R
+a_{11} s^\la_{\pt{\la}\la} s^\mu_{\pt{\mu}\mu} R 
\Big].\nonumber\\
\label{dynS}
\eea
The first two terms are just the originally proposed SME couplings, 
linear in the coefficients $s_\mn$.
The remaining terms are second order in 
the coefficients $s_\mn$ \cite{b19}. 
Since $s_\mn$ are dimensionless and normally assumed small
compared to unity, 
these terms represent a step beyond the minimal SME, 
which assumes first-order terms in the coefficients, 
and they are a special case of the terms outlined in Table V and Table VII in Ref.\ \cite{kl20}.

Many of these terms for a symmetric two-tensor 
have been proposed in modified gravity models 
in the literature in different contexts \cite{cmw}.
Also, other possible terms are omitted due to equivalence via integration by parts.
For example,
\bea
0 &=& \int d^4x  
\sqrt{-g} \big( \nabla_\ga s_{\al\be} \nabla^\be s^{\al\ga} 
- \nabla_\be s_\al^{\pt{\al}\be} \nabla_\ga s^{\al\ga} 
\nonumber\\
&&
\pt{\int d^4x}
-s_{\al\be} s_{\ga\de} R^{\al\de\be\ga}
+s_{\al\de} s^\al_{\pt{\al}\be} R^{\de\be} \big).
\label{ibp}
\eea

Note also that one can add general potential terms for a symmetric two-tensor 
of the form $V(s_\mu^{\pt{\mu}\mu},s_\mn s^\mn, \hdots)$ for the case of spontaneous symmetry breaking, 
as detailed elsewhere \cite{kp09}.
In the particular case of the match to 3+1 models in section \ref{horavamatch}, 
the possibility exists of using a term quartic in the coefficients $s_\mn$:
\beq
\Delta \cL_s = a_{12} \fr{\sqrt{-g}}{2\ka} (s^\mn \nabla_\mu s_{\nu\la}) 
(s^{\ka\rh} \nabla_\ka s_\rh^{\pt{\rh}\la}).   
\label{quartic}
\eeq
An analysis of these and other possible terms in the SME is forthcoming.


\end{document}